\documentclass[fleqn,usenatbib, useAMS]{mnras}
\usepackage{subfig}
\usepackage{subcaption}
\hypersetup{
    colorlinks=true,
    linkcolor=blue,
    filecolor=magenta,      
    urlcolor=cyan,
    pdftitle={Overleaf Example},
    pdfpagemode=FullScreen,
    }
\usepackage[normalem]{ulem}
\usepackage{booktabs}
\usepackage{amssymb}	
\usepackage{placeins}
\usepackage{newtxtext,newtxmath}

\usepackage[T1]{fontenc}

\newcommand{\sqcm}{cm$^{-2}$\,}  
\newcommand{\kms}{$\rm km\, s^{-1}$\,} 
\newcommand{\lya}{Ly$\alpha$\,}
\newcommand{\lyb}{Ly$\beta$}

\newcommand{\HI}{\mbox{H\,{\sc i}}}

\newcommand{\CIV}{\mbox{C\,{\sc iv}}}

\newcommand{\CII}{\mbox{C\,{\sc ii}}}

\newcommand{\CIII}{\mbox{C\,{\sc iii}}}

\newcommand{\SiII}{\mbox{Si\,{\sc ii}}}
\newcommand{\SiIII}{\mbox{Si\,{\sc iii}}}
\newcommand{\SiIV}{\mbox{Si\,{\sc iv}}}

\newcommand{\MgI}{\mbox{Mg\,{\sc i}}}
\newcommand{\MgII}{\mbox{Mg\,{\sc ii}}}
\newcommand{\FeII}{\mbox{Fe\,{\sc ii}}}

\newcommand{\AlII}{\mbox{Al\,{\sc ii}}}
\newcommand{\AlIII}{\mbox{Al\,{\sc iii}}}

\newcommand{\logN}{$\log_{10} N/{\rm cm^{-2}}$} 
\newcommand{\Msun}{M$_{\odot}$}
\newcommand{\ang}{\mbox{\normalfont\AA}}
\newcommand{\NHI}{$N({\rm H \textsc{i}}$)}

\newcommand{\nh}{$n_{\rm H}$}
\newcommand{\met}{$[X/H]$}
\newcommand{\logz}{$\log{Z/Z_{\odot}}$}

\newcommand{\lum}{$L_{\rm Ly\alpha}$}

\DeclareRobustCommand{\VAN}[3]{#2}
\let\VANthebibliography\thebibliography
\def\thebibliography{\DeclareRobustCommand{\VAN}[3]{##3}\VANthebibliography}

\usepackage{graphicx}	
\usepackage{amsmath}	

\title[MUSEQuBES: physical properties of the gas around LAEs]{MUSEQuBES: Investigating the Physical and Chemical Properties of Circumgalactic Gas Around Ly$\alpha$ Emitters at $z \approx 3.3$}

\author[E. Banerjee et al.]{
Eshita Banerjee $^{1}$\thanks{E-mail: eshitaban18@iucaa.in},
Sowgat Muzahid $^{1}$,
Joop Schaye $^{2}$,
Sean D. Johnson $^{3}$,
and Sebastiano Cantalupo $^{4}$
\\ $^{1}$ IUCAA, Post Bag 04, Ganeshkhind, Pune-411007, India\\
$^{2}$ Leiden Observatory, Leiden University, P.O. Box 9513, NL-2300 AA Leiden, the Netherlands\\
${^3}$ Department of Astronomy, University of Michigan, 1085 S. University, Ann Arbor, MI 48109, USA\\
${^4}$ Department of Physics, University of Milan Bicocca, Piazza della Scienza 3, I-20126 Milano, Italy
}

\date{Accepted XXX. Received YYY; in original form ZZZ}

\pubyear{\the\year{}}

\begin{document}
\label{firstpage}
\pagerange{\pageref{firstpage}--\pageref{lastpage}}
\maketitle

\begin{abstract}
\noindent{We present a detailed study of the physical properties of absorbers associated with $z\approx3.3$ \lya\ emitters (LAEs) from the MUSEQuBES survey. Using \HI\ and ionic column densities derived from Voigt profile fitting, we determine the density and metallicity of 75 absorbers associated with 59 LAEs through a custom-built Bayesian framework. The overall sample shows a median density $\log\,(n_{\rm H}/{\rm cm^{-3}})=-2.7\pm0.7$ and metallicity \met$=-1.6^{+1.2}_{-0.9}$ with $\approx15\%$ ($\approx 40\%$ including upper limits) of systems showing metallicities consistent with the IGM at these redshifts (\met$\lesssim -2.8$). Intriguingly, absorbers with $\log\, N(\rm HI)/{\rm cm^{-2}} \gtrsim 16.5$, corresponding to an overdensity of $\sim 100$ at $z=3$, exhibit a bimodal metallicity distribution with peaks at \met$=-3.8\pm 0.2$ and $-1.8\pm 0.6$. The latter are observed at large impact parameters ($\gtrsim150$~pkpc) and often exhibit low-ionization species (e.g., \SiII, \AlII). We interpret the former as pristine inflowing gas from the cosmic web, while the latter likely traces metal-enriched CGM associated with outflows from faint galaxies not detected in Ly$\alpha$ emission. We find no significant correlations between absorber metallicity or density and host galaxy properties, including redshift, impact parameter, SFR, Ly$\alpha$ luminosity, or environment. Absorbers separated by $\lesssim500$~\kms\ show $\sim$1~dex metallicity and $\sim$0.5~dex density variations, indicating physical and chemical inhomogeneity of the medium around these LAEs.}

\end{abstract}






\section{INTRODUCTION}

The distribution of elements heavier than hydrogen and helium, collectively referred to as metals, serves as a fossil record of star formation and feedback across different cosmic epochs. Formed in the interiors of stars through nucleosynthesis, these metals are expelled into the surrounding medium via stellar winds, supernovae, and large-scale galactic outflows. At low redshift, estimates suggest that stars and the interstellar medium (ISM) together contain no more than $\approx20\%$ of the heavy elements produced in stellar interiors, indicating that the bulk of the metals are ``missing'' from galaxies \citep{Peeples_2014}.


These missing metals are likely have mixed with the ambient medium around galaxies, enriching both the circumgalactic medium \citep[CGM;][]{Tumlinson2017, Chen2024} and the more extended intergalactic medium (IGM). These extended environments act not only as reservoirs for the majority of the universe’s metal content, but also as dynamic sites of interaction, where accretion and feedback processes regulate the flow of matter and energy. In particular, the CGM plays a critical role in galaxy evolution by regulating gas cooling, star formation, and metal recycling. Mapping the distribution of metals in and around galaxies across redshift is therefore key to understanding the baryon cycle and the underlying physical processes that shape galaxies \citep[e.g.,][]{Peroux2020}.

{The epoch around $z \approx 2$–3, often referred to as ``cosmic noon'', marks the peak of both cosmic star formation and black hole growth \citep{Madau2014review}. To trace how the enriched CGM and IGM that fueled this activity were assembled, it is essential to look further back in time. The redshift interval $z \approx 3$–4 offers a crucial window into this formative stage, when early gas accretion and feedback were actively shaping the halos of galaxies.}

Although emission-line studies offer a direct way to trace metals in galaxy halos, they are often hindered by the inherently low surface brightness of diffuse gas. A major breakthrough enabled by modern integral field spectrographs such as VLT/MUSE \citep{bacon2010muse} is the direct detection of metal-line emission from extended gaseous halos. Such emission has now been observed in diverse environments, around quasars \citep{Borisova_2016,Johnson2018, Fabrizio2019, Johnson_2022}, star-burst galaxies \citep{Rubin2011, Rupke2019, Burchett2021}, and galaxy groups or clusters \citep{Chen2019, Leclercq2022}. However, detecting such metals in emission around normal galaxies has so far required stacking large samples to reach the necessary sensitivity \citep[e.g.,][]{Erb_2012,Guo2023Natur,Dutta2023MgII_emission, Kusakabe2024SiII}, which limits our ability to examine individual systems in detail.



In contrast, absorption-line spectroscopy has long proven to be a more sensitive tool for probing diffuse, metal-enriched gas. By analyzing absorption features imprinted on the spectra of background quasars or galaxies, we can detect metals even in regions with densities far lower for emission line studies. This technique reveals the presence of heavy elements at higher redshifts from the tenuous IGM, as seen in the metal-enriched, low-density \lya\ forest \citep[e.g.,][]{Schaye_2003, Simcoe2004, Aguirre2008, muzahid2012high, kodiaq_z}, to the higher density Lyman Limit Systems \citep[LLSs; e.g., ][]{Prochaska2006, Fumagalli2013, Fumagalli2016_fila, Banerjee2025_filament} and damped \lya\ absorbers \citep[DLAs; e.g., ][]{Ledoux2006, Noterdaeme2009, Fynbo_2010, Peroux_2012, Mackenzie_2019}.

Some of the initial attempts to measure the metallicity of individual high-redshift absorbers date back to \citet{Steidel1990}, who analyzed a sample of eight absorption systems in the redshift range $2.90 < z < 3.23$ with \HI\ column densities in the range $17.0 < \log (N_{\rm HI}/\rm cm^{-2}) \lesssim 19.3$ (i.e., LLSs). Using photoionization modeling, they estimated heavy element abundances spanning from $\log\, Z/Z_{\odot} \approx -3.0$ to $-1.5$. This pioneering study opened a window into the metal content of the high-$z$ universe.

Subsequent high-$z$ studies have primarily targeted {a handful of} $N({\rm HI})\gtrsim 10^{19}$~\sqcm\ absorbers, since their damping wings enable precise \HI\ column density measurements. At $z<2$, a handful of sub-DLA ($19 \lesssim \log (N_{\rm HI}/ \rm cm^{-2}) \lesssim 20.2$) studies suggested that these systems are more metal-rich (typically of solar metallicity) than DLAs \citep[$\approx 1/10^{\rm th}$ of solar metallicity; e.g., see ][]{ Kulkarni2007, Peroux2008}. Later, \citet{Fumagalli2016} presented metallicities of a substantially larger sample of LLSs at $z > 2$ and found that these systems are notably metal-poor, with a mean metallicity around $\log\, Z/Z_{\odot} \approx -2$, significantly lower than that of DLAs. The recent KODIAQ-Z survey examined a wide range of absorber strengths ($14.6 < \log\, (N_{\rm HI}/ \rm cm^{-2}) < 20$) at $2.2 < z < 3.6$ \citep{kodiaq_z}. On average, they found the metallicity of these \HI\ absorbers to be $\approx -2.2$, nearly 1 dex lower than that of their $z < 1$ counterparts, which have an average metallicity of $\approx -1.2$.

More recently, the MAGG survey examined a sample of more than 50 LLSs and DLAs detected within $\pm1000$~\kms\ (and $\approx 300$~physical kpc; hereafter, pkpc) of $z=3$–4 \lya\ emitters \citep[LAEs, see][]{Lofthouse_2023}. They reported a median metallicity of \met$=-2.5 \pm 0.5$, with nearly $20\%$ of the systems being extremely metal-poor (\met$<-3$). Similar to other absorber-centric studies\footnote{Absorber-centric studies start with the absorber redshifts and identify galaxies associated with them.}, \citep[see also,][]{Fumagalli2016_fila, Mackenzie_2019, Lofthouse2020}, they argued that such absorbers likely trace inflowing gas along cosmic filaments connecting galaxies, though the filaments themselves were not directly observed. A direct detection was recently reported by \citet{Banerjee2025_filament}, who identified a filamentary structure in \lya\ emission connecting seven LAEs at $z=3.6$. The associated absorber has an exceptionally low metallicity (\met$=-3.7$), consistent with a primordial origin, thereby linking the galaxy overdensity and the metal-poor gas to the same underlying cosmic web. Nevertheless, these studies present relatively strong absorbers (LLSs and DLAs), leaving the metallicity of weaker \HI\ systems associated with galaxies at these redshifts largely unconstrained.

%

In this study, we investigate the physical properties of the absorbers surrounding the LAEs at $z\approx3$. Metal enriched gas in the CGM is expected to eventually accrete onto galaxies \citep[e.g.,][]{Turner_2017}, fueling future star formation and contributing to the enhanced star formation rate density observed at $z\approx1$–2. Previously, \cite{Zahedy2019} tried to constrain the metallicity of the absorbers associated to few LAEs at $z\sim3$ using deep long-slit observations and found the trend of large scale gas accretion. In this study, we make use of the data from the MUSE Quasar-fields Blind Emitters Survey \citep[MUSEQuBES; ][]{Muzahid_2020, Muzahid_2021}, which combines deep integral field spectroscopy from MUSE with high-resolution quasar spectra from VLT/UVES and Keck/HIRES. MUSEQuBES specifically targets absorbers located within a few hundred pkpc (median $\approx 160$~pkpc) of LAEs at $z \approx 3.3$. As demonstrated in our earlier work, the majority of these absorbers span an \HI\ column density range of $15 \lesssim \log (N_{\rm HI}/ \rm cm^{-2}) \lesssim 17$ \citep{Banerjee2025_HI}. In this paper, we adopt a galaxy-centric approach to investigate the metallicity and density of the absorbers associated with the LAEs, without imposing any prior selection on their \HI\ strength. 



This paper is organized as follows: Section~\ref{sec:data} describes the data analyzed in Section~\ref{sec:analysis}. The main results are presented in Section~\ref{sec:results}. Section~\ref{sec:Discussions} provides a discussion of these results followed by the conclusion in Section~\ref{sec:Conclussion}. 
We adopt a flat $\Lambda$CDM cosmology with $H_0$ = $70\,\rm km\, s^{-1}\, Mpc^{-1}$, $\Omega_{\rm M} = 0.3$ and $\Omega_\Lambda = 0.7$. Metallicity ($\equiv$\met) is expressed on a logarithmic scale relative to solar metallicity \citep[$\rm Z_{\odot}=0.013$; ][]{Asplund2009}. Distances are in physical kpc unless specified otherwise.


\section{Data}
\label{sec:data}

The quasar and galaxy data used in this study are drawn from the high-redshift part of the MUSEQuBES survey \citep{Muzahid_2020, Muzahid_2021, Banerjee_2023, Banerjee2025_filament, Banerjee2025_HI}\footnote{For the low-$z$ MUSEQuBES survey, the readers are referred to \citet{Dutta2024,Dutta2025a,Dutta2025b}.}. The high-$z$ MUSEQuBES includes 8 MUSE fields, each centered on a UV-bright quasar ($3.9 \gtrsim z_{\rm qso}\gtrsim3.5$). It is designed to study the CGM of high-$z$ ($z\approx3.3$) LAEs using absorption lines seen in the quasar spectra. The following sections briefly describe the data used in this study.

\subsection{Quasar spectra}

The optical spectra of the quasars, with a spectral resolution of $R \approx 45{,}000$, were obtained from the SQUAD \citep[VLT/UVES;][]{Murphy2019} and the KODIAQ databases \citep[Keck/HIRES;][]{O_Meara_2015}. These spectra have a median signal-to-noise ratio (SNR) of $\gtrsim 30$ in the \lya\ forest and $\gtrsim 70$ redward of the \lya\ emission \citep[see Table 1 from][for further details about the observations]{Muzahid_2021}.

We further use the near-infrared data of these quasars from VLT/X-shooter, covering 1000--2480~nm with a spectral resolution of $ \approx 5300$. The median SNRs of these spectra fall in the range $\approx30-70$~ per pixel, corresponding to $3\sigma$ detection limits for absorption features with equivalent widths of $\approx0.02$–$0.01$~\ang, computed over 100~\kms\ velocity window i.e., roughly twice of the spectral resolution. For four of the eight fields (BRI1108$-$07, J0124$+$0044, Q0055$-$269, Q1317$-$0507), the best-fitting continua were obtained from the ESO archive \citep{Lopez2016}. For the remainder, we used reduced but un-normalized spectra from the ESO archive \citep{Galbiati2024}, and fitted the local continuum with a custom script. The script bins the spectra into windows of a few tens of pixels, and computes the running median flux in each bin. This approach provides a robust estimate of the underlying spectral shape even in the presence of strong absorption or emission features. The resulting median points were interpolated using the {\sc scipy} interpolation function and further smoothed with the 1D convolution function from {\sc astropy} to remove small-scale fluctuations. The final continuum model is suitable for spectrum normalization. Although the resolution of X-shooter spectra is too low for Voigt profile fitting, they still provide useful upper limits on column densities for un-detected lines.

\subsection{MUSE data and the LAE sample} 
\label{sec:LAE_data}

A total of 50~hours of guaranteed time MUSE observations (PI: J. Schaye) across 8 quasar fields allowed us to identify 96 foreground LAEs with median \lya\ luminosity $\approx 10^{42}\, \rm erg\, s^{-1}$ at $2.9 < z < 3.8$ detected via their \lya\ emission. These LAEs span an impact parameter range of $16<\rho<320$~pkpc. The \lya\ line often shows velocity offsets of several hundred \kms\ from the systemic redshifts of the galaxies \citep[e.g., ][]{Steidel_2010, Rakic+11, Shibuya_2014, Verhamme18}. To estimate the systemic redshifts, we used an empirical relation between this velocity offset and the Full Width at Half Maximum (FWHM) of the \lya\ emission peak, derived for our sample by \cite{Muzahid_2020}:
$V_{\rm offset} = 0.89\times {\rm FWHM} -58$~\kms.

The star formation rates (SFRs) of these LAEs are derived from the UV luminosity density using the calibration of \cite{Kennicutt_1998}, adjusted to the \cite{Chabrier_2003} initial mass function. For 39 out of the 96 LAEs, we could measure the SFR from their UV continuum. For the rest, the UV continuum was either undetected at $>5\sigma$ confidence or contaminated by foreground sources \citep[see ][]{Muzahid_2020}. For the cases of undetected continuum, we report SFR upper limits. The median SFR for this sample (detections only) is $\approx 1.2$~\Msun~$\rm yr^{-1}$. Note that the SFRs are not corrected for dust extinction which is expected to be low for such low-mass galaxies.

Stellar masses ($M_*$) were estimated from the SFRs using the star-forming main sequence relation from \citet{Behroozi2019}. The median stellar mass of our sample is $M_* = 10^{8.9}$~\Msun. While this method involves some assumptions, the values are consistent with typical LAEs at these redshifts, which have $M_* = 10^8 - 10^{8.5}$~\Msun\ and SFR $\approx 2.3$~\Msun$/\rm year$, based on SED fitting and rest-optical emission line studies \citep[see,][]{Ono2010, Trainor2015}. However, their \lya\ line luminosities are somewhat higher than those of our sample. Although we do not have an estimate of dust extinction for these LAEs, previous multi-band studies have shown that for the LAEs at $z\sim2$, the typical values are $A_v=0.12^{+0.25}_{-0.08}$ \citep{Hao2018}. For detailed information on the MUSE data analysis and the properties of LAEs, we refer the reader to \citet{Muzahid_2021}.

\section{Analysis}
\label{sec:analysis}

\subsection{General information about the measurements of absorption profiles}
\label{sec:abs_measurment}

The Voigt profile decomposition of all \CIV\ absorbers associated with the LAEs in the MUSEQuBES sample has already been presented by \cite{Banerjee_2023}. Here, we extend the Voigt profile analysis to additional metal transitions, including \CII, \SiII, \AlII, \FeII, \CIII, \SiIII, \AlIII, and \SiIV, using the high-resolution UVES and/or HIRES spectra. For non-detections of metal transitions, we estimate $3\sigma$ upper limits on column densities from the $3\sigma$ limiting equivalent width \citep[EW;][]{Hellsten1998}, assuming the linear portion of the curve of growth (CoG). We further used X-shooter spectra to examine \MgI\ and \MgII\ transitions; however, owing to the lower spectral resolution, Voigt profile fitting was not feasible. Consequently, these lines were excluded from the modeling when detected \footnote{Note that, \MgII\ is detected for only 4 systems in our sample, 2 of them being DLAs.}. However, in cases of non-detections, we used their column density upper limits derived from the spectra.

Metal abundances must be measured relative to hydrogen, and hence we also incorporated the \HI\ absorber catalog from \citet{Banerjee_2023}\footnote{Note that, during this work, some of the \CIV\ and \HI\ component parameters from our previous catalog were slightly adjusted when fitting the additional metal transitions, as constraints from aligned metals were not imposed in \citet{Banerjee2025_HI}. In a few cases, weak low-column density components were also added. These updates, however, have a negligible impact on our previous results, as they were based on the total column densities which remain unaffected by such weaker components.}. In that study, we accounted for all available higher-order Lyman-series lines simultaneously, as accurate determination of \HI\ column densities in high-$z$ quasar spectra is often hindered by line saturation and/ or blending within the \lya\ forest. This simultaneous fitting approach not only improves the accuracy of \HI\ column density measurements but also helps to identify and mitigate contamination. When no clean, unsaturated higher-order lines were available, we reported the \HI\ column density as a lower limit for those components.

Hereafter, in this study, we adopt the following convention: an absorption {\em component} refers to an individual Voigt profile, whereas a {\em system} (or absorber) represents one or more components grouped together in velocity space. For ionization modeling, we use the total column density of each ion measured in individual system.

The grouping of components into systems followed the following approach. For each LAE, we begin with the \CIV\ components detected within $\pm500$~\kms. Next we apply a one-dimensional friends-of-friends algorithm with a linking velocity of $40$~\kms\ in order to define a system. This choice of linking velocity is motivated by the small-scale clustering length scale measured for the blind \CIV\ absorber catalog presented in \cite{Banerjee_2023}.

Establishing a one-to-one correspondence between metal-line components (e.g., \CIV) and \HI\ components is not straightforward. This is because (a) \HI\ lines have larger Doppler parameters due to the smaller mass of hydrogen, and (b) they exhibit higher optical depths owing to both the high cosmic abundance of hydrogen and the large $f\lambda$ values ($f$ being the oscillator strength and $\lambda$ is the rest frame wavelength of the transition) of the lower-order \HI\ Lyman-series transitions. In addition, the higher-order Lyman lines are either not covered in the data or are severely blended within the \lya forest. Therefore, instead of attempting a direct component-wise association, we grouped all \HI\ absorbers detected within $z_{\rm min} - 40$~\kms\ to $z_{\rm max} + 40$~\kms, where $z_{\rm min}$, $z_{\rm max}$ are the minimum and maximum redshift of metal absorption components (most often \CIV) of a system. Fig.~\ref{fig:example_veloplot} illustrate how components closely separated in velocity space are grouped into multiple systems.

Out of the 96 LAEs in our sample, 37 show no detectable metal transitions within $\pm500$~\kms\ of their corrected redshifts and therefore they are excluded from the further analysis. For the remaining 59 LAEs, we identified 75 unique absorption systems, which we could further consider for ionization modeling. \CIV\ is most commonly detected transition in these systems, except for one LAE along the PKS1937$-$101 sightline, where \CIV\ falls outside the spectral coverage. 

For each system, we searched for all available aligned metal transitions. When \CIV\ is detected, the velocity structures of the other ions generally closely follow that of \CIV, suggesting a common gas phase; in such cases, their redshifts were tied to the corresponding \CIV\ components during the fitting. out of the 75 systems, 30 exhibit secure \SiIV\ detections. For 16 systems, 3$\sigma$ upper limits on $N($\SiIV$)$ could be derived, while in the remaining 29 cases, \SiIV\ is contaminated and therefore excluded from the modeling. Similarly, \CIII\ and \SiIII\ are almost always heavily blended with the \lya\ forest and are also excluded from the analysis.

We have further identified 11 systems where low-ionization transitions e.g., \SiII, \AlII; are present; and they are predominantly associated with the higher \NHI\ absorbers. Two of these systems show no detectable high-ion absorption despite complete spectral coverage of both \CIV\ and \SiIV. For the remaining nine systems, where both high-ionization transitions (e.g., \CIV, \SiIV; hereafter ``high ions'') and low-ionization transitions (e.g., \SiII, \AlII; hereafter ``low ions'') are present, the two phases were fitted independently. Among these, we identify four systems in which the low and high-ion absorption exhibits significantly different component structures; all have $\log\, N(\rm HI)/{\rm cm^{-2}} > 17.2$. The two DLAs in our sample fall into this category, and their best-fitting Voigt profiles are shown in Fig.~\ref{fig:DLA_vpfit}. By contrast, Fig.~\ref{fig:no_dif_low_n_high_ion_veloplot} presents a system in which both low and high-ions are detected and exhibit similar kinematic structures. Additionally, the \CII\ lines are generally contaminated by the \lya\ forest, except in 17 systems where the \CII$\lambda1334$ transition falls redward of the forest; among these, \CII\ is detected in two cases. 
 

\subsection{Photoionization modeling} 
\label{sec:Photoionization_model}

\begin{table}
    \centering
    \caption{Range of input parameters used for generating {\tt Cloudy} grids} 
    \begin{tabular}{cccc}
    \hline
    Parameter & Minimum & Maximum  & Interval \\
    \hline
    $\log N_{\rm HI}/\rm cm^{-2}$ & 12.5 & 20.5 & 0.25 \\
    $z$  & 2.75 & 4 & 0.25 \\
    \met & $-$4.0 & 1.0 & 0.25 \\
    $\log n_{\rm H}/\rm cm^{-3}$  & $-$4.5 & 0.0 & 0.25 \\
    \hline
    \end{tabular}
    \label{tab:cloudy_params}
    \vspace{.2cm}\\ Note:-- {\met\ is expressed as \logz.} 
\end{table}
%


Our photoionization modeling follows an approach similar to \citet{Fumagalli2011, Crighton2015, CCCI}, utilizing the {\tt Cloudy} code (version C17; \citealp{Ferland2013}) to simulate a uniform slab of gas in photoionization equilibrium. The ionizing radiation includes the redshift-dependent UV background from \citet[][hereafter HM05]{Haardt_Madau_2001} and the cosmic microwave background. We assume a constant hydrogen density (\nh), corresponding to a single-phase medium. Elemental abundances follow solar values from \citet{Asplund2009}. Each model runs until the specified \NHI\ is reached, setting the stopping criteria for {\tt Cloudy.} We assume all elements are in the gas phase and neglect any depletion onto dust grains.

The model grid spans the following parameters: (i) neutral hydrogen column density (\NHI), (ii) redshift ($z$), (iii) metallicity (\met), and (iv) the total hydrogen number density (both neutral and ionized: \nh), which is related to the ionization parameter ($U$) as: $U = n_{\gamma}/n_{\rm H}$; $n_{\gamma}$ being the hydrogen ionizing photon density. The parameter ranges are summarized in Table~\ref{tab:cloudy_params}. We generated the model grid varying each parameter in steps of 0.25 dex. Densities and metallicities were derived using a Bayesian approach, by comparing the measured column densities (including limits) of each absorber against this grid. In order to get the intermediate values during sampling, we applied linear interpolation\footnote{We have used \href{https://docs.scipy.org/doc/scipy/reference/generated/scipy.interpolate.RegularGridInterpolator.html}{RegularGridInterpolator} from Scipy.}.

{To compute the posterior probability distributions for the model parameters, we used the likelihood functions described in \cite{Fumagalli2016}, which takes the observed column densities and their uncertainties for different ions as input. Here to mention, for non-detections, we adopted an uncertainty of $\Delta \log N_{\rm lim} = 0.14$\footnote{Assuming that the $3\sigma$ upper limits correspond to $1\sigma$ errors, i.e., $\Delta \log N_{\rm lim} = \frac{1}{{\rm ln}\, 10} \cdot \frac{\sigma}{3\sigma}$}. Using these likelihood functions, the parameter space was explored using the nested sampling algorithm \texttt{MLFriends}, implemented in the {\tt UltraNest}\footnote{\url{https://johannesbuchner.github.io/UltraNest/}} Python package \citep{Buchner_2021}. Since \NHI\ and $z$ are tightly constrained by {\tt Vpfit}, we adopted Gaussian priors on these parameters with widths set by their respective measurement uncertainties returned by {\tt Vpfit}. Our only deviation from the approach of \cite{Fumagalli2016} arises for absorbers where \NHI\ is reported as a lower or upper limit: in these cases, we still applied Gaussian priors on \NHI\ with a large sigma ($\approx 1~dex$) during sampling, but interpreted the resulting metallicities as upper/ lower limits. For the remaining model parameters, metallicity and \nh, we used flat priors spanning the full extent of our model grid.}

Exceptions were made for \nh\ priors in systems with limited observational constraints, typically those with only one detected metal line (most often \CIV). In such cases, we adopted Gaussian priors on \nh\ with a mean value of $\log_{10}($\nh$/{\rm cm^{-3}})\approx -3.5$ and a standard deviation of $1$~dex. This is motivated by the fact that, for absorbers with \NHI~$<10^{17}$~\sqcm\ at $z \approx 3-4$, the \CIV\ ion fraction peaks at $\log_{10}($\nh$/{\rm cm^{-3}})\approx -3.5$, which is largely independent of metallicity. Systems with \NHI$>10^{17}$\sqcm\ usually exhibit multiple detected metal transitions and could therefore be modeled with flat \nh\ priors alone. Fig.~\ref{fig:example_model} show examples of our {\tt Cloudy} modeling exercise.

{Finally, note that, we have not explored potential non-solar abundance patterns in these absorbers, as their densities are primarily constrained by \CIV\ and \SiIV, with no additional metal transitions commonly detected.}

\section{Results} 
\label{sec:results}

\subsection{Density of the absorbers}
\label{subsec:hden}

\begin{figure}
    \centering
    \includegraphics[width=\linewidth]{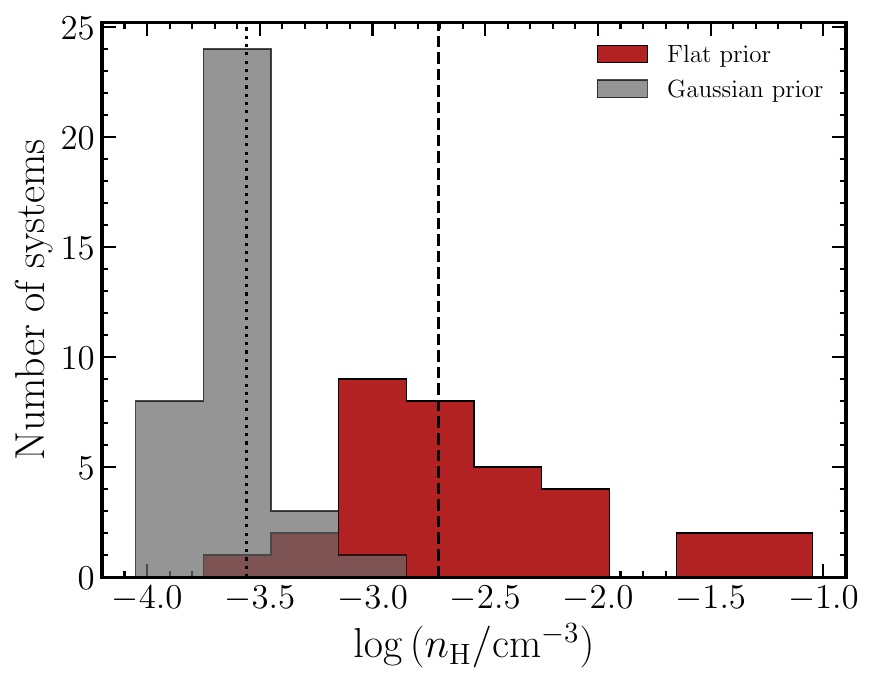}
    \caption{Distribution of \nh\ for the absorbers: maroon denotes values estimated using a flat prior in the Bayesian sampling, while gray corresponds to those with a Gaussian prior of $\log\,($\nh$/{\rm cm^{-3}})= -3.5\pm 1.0$. The median \nh\ of the systems modeled by a flat (Gaussian) prior is indicated by the vertical dashed (dotted) line.}
    \label{fig:hist_hden}
\end{figure}

In this section, we examine the total hydrogen number density (\nh) of the absorption systems. As mentioned above, systems with only \CIV\ detections were modeled using a Gaussian prior on \nh. For the 2 DLAs in our sample, we could not constrain the density, since at such high \NHI\ values, the ion fractions of low- and intermediate-ionization lines are insensitive to density. The density distributions of the systems, obtained assuming flat and Gaussian priors, are shown by the red and grey histograms in Fig.~\ref{fig:hist_hden}. A total of 36 systems were modeled with a flat prior, while 37 systems were modeled with a Gaussian prior. For the systems where \nh\ is derived using a flat-prior, the mean hydrogen number density is $\log\,(n_{\rm H}/{\rm cm^{-3}}) = -2.7$ (same as median) with a standard deviation of $0.7$. For the  systems where \nh\ is derived using a Gaussian-prior, both the median and mean are $\log\,(n_{\rm H}/{\rm cm^{-3}}) = -3.6$ with a standard deviation of $0.2$. 
The offset between the two distributions likely arises from the different constraints used in the two modeling approaches. In the Gaussian-prior cases, \nh\ is driven toward the density where the \CIV\ ion fraction peaks, which generally corresponds to lower densities. In contrast, the flat-prior cases are further constrained by transitions with lower ionization potentials (e.g., \SiII, \SiIV) than \CIV, pushing the preferred solution toward higher densities.



We next investigate whether \nh\ correlates with any of the galaxy properties. Here we restrict the analysis to the 36 systems for which \nh\ could be constrained using flat priors. First, we do not observe any significant redshift evolution of \nh\ ($p$-value from a Spearman correlation test = 0.3). Among the 36 systems, 11 are associated with LAEs in pairs/ groups, while 25 are linked to isolated LAEs. The \nh\ distributions of these two subsamples are not statistically different, as confirmed by a KS-test ($p$-value = 0.4). 

By default, we associate every absorber within $\pm500$~\kms\ of an LAE to that galaxy. When the same absorber meets this criterion for multiple LAEs, we adopt three association methods: (a) linking the absorber to all such LAEs, which means it is counted multiple times, once for each galaxy; (b) assigning the absorber to the LAE with the smallest projected separation (impact parameter); and (c) to the LAE with the smallest three-dimensional distance, $r_{\rm 3D} = \sqrt{r_\perp^2 + r_\parallel^2}$, where $r_\parallel$ is the line-of-sight velocity separation converted to physical distance (assuming pure Hubble flow) at the absorber's redshift, defined as the \NHI-weighted mean redshift of the individual \HI\ components derived from {\tt Vpfit}. Method (a) is adopted as our fiducial approach in the main analysis, while the results based on methods (b) and (c) are presented in Appendix~\ref{sec:appendix_galaxy_param}. The key conclusions remain consistent across all three approaches.

\begin{figure*}
    \centering
    \hbox{
    \includegraphics[width=0.35\linewidth]{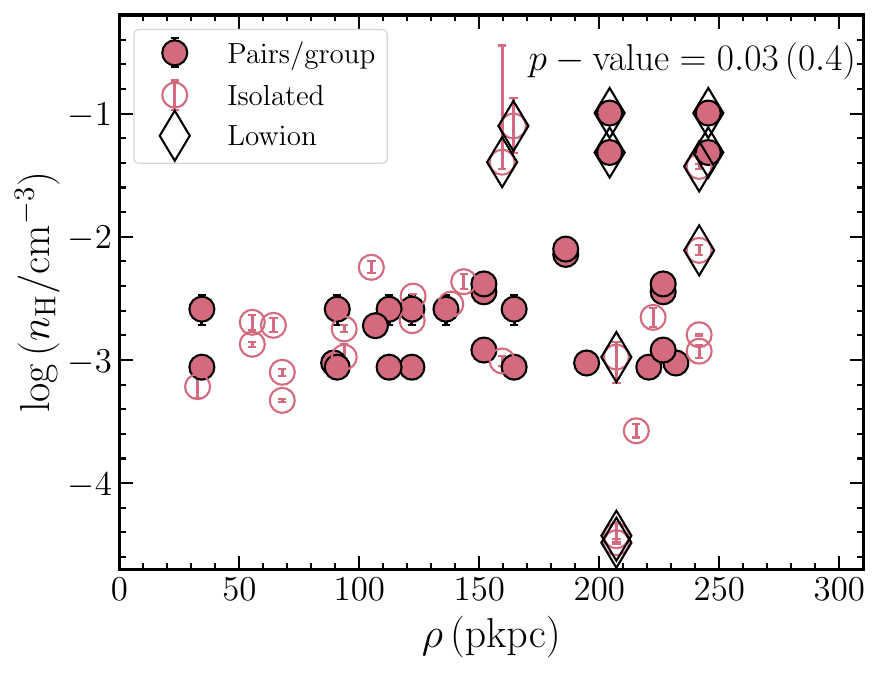}
    \includegraphics[width=0.3\linewidth]{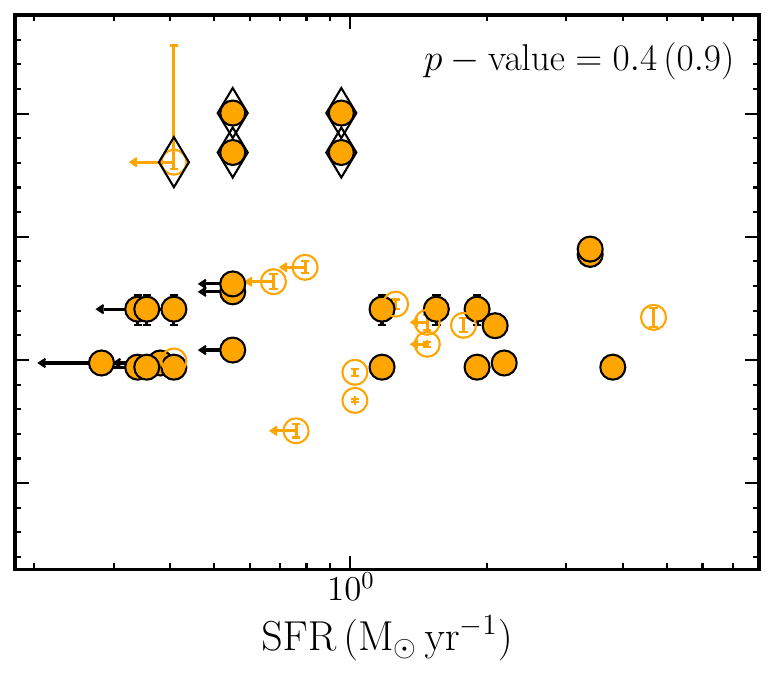}
    \includegraphics[width=0.3\linewidth]{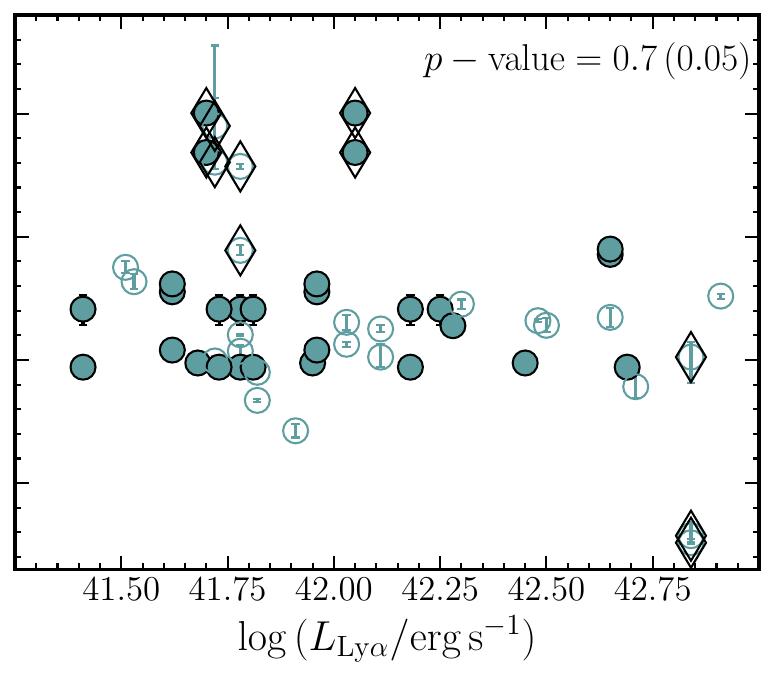}
    }
    \caption{The total hydrogen number density as a function of impact parameter {\tt (left)}, SFR {\tt (middle)}, and \lya\ line luminosity {\tt(right)}. Only systems where \nh\ could be constrained using flat priors are considered here. The $p$-values from the Kendall-$\tau$ correlation tests are reported outside (within) parentheses for group (isolated) LAEs. {The results indicate a strong correlation between \nh\ and impact parameter for the group subsample, and anti-correlation between \nh\ and luminosity for the isolated subsample.} However, these trends are largely driven by systems with detected low-ion transitions (shown as diamonds). This is further discussed in the main text.}
    \label{fig:hden_params}
\end{figure*}

Fig.~\ref{fig:hden_params} shows \nh\ as a function of impact parameter, SFR, and \lya\ luminosity of the associated LAEs, with empty (filled) circles marking isolated (group) systems. The left panel reveals a strong correlation between \nh\ and impact parameter ($\rho$) of the pairs/ group LAEs, as indicated by the Spearman rank correlation test shown in the plot ($p=$0.03). However, this trend is largely driven by systems with detected low-ion absorption (marked by diamonds), all of which have $\rho>150$~pkpc. As noted earlier, the low-ions are detected in higher \NHI\ systems. Thus, the observed trend could arise from the fact that the true host of these systems remained undetected via our \lya\ selections; an issue discussed further in Section~\ref{sec:met_discuss}. Excluding these systems, the correlation between \nh\ and $\rho$ largely disappears ($\rho_{\rm Sp}=0.3$, $p=0.08$). The middle and right panels plot \nh\ against SFR and \lya\ luminosity (\lum), respectively. {An apparent (anti-)correlation between \nh\ and \lum\ is observed for the isolated galaxies, which too is largely driven by systems with detected low ions.}


\subsection{Metallicity measurements} 

In this section, we present the metallicity measurements for the 75 systems introduced in Section~\ref{sec:abs_measurment}. Two of these are classified as DLAs based on the \NHI\ values. Since ionization corrections are negligible for such high-\NHI\ systems, their metallicities were determined directly from the \SiII\ column densities. In addition to these, we identify 9 non-DLA systems where low-ion transitions are detected. As discussed in Appendix~\ref{sec:appendix_toy_model}, whenever low ions are present, the metallicity is constrained solely by them, without taking the high-ionization lines into consideration. For the remaining 64 systems, metallicity estimates rely on the measured column densities of high ions.

We have 23 systems for which \NHI\ is reported only as a lower limit, therefore, their metallicity is treated as an upper limit (i.e., ``Uplims''). Additionally, as discussed in Section~\ref{sec:abs_measurment}, our method of associating \HI\ and metal lines can result in two consecutive systems sharing the same \HI\ component. In such cases, the \NHI\ of that component is treated as the maximum value that can be associated among the shared components. Since our modeling depends on the total \NHI, if the contribution from the shared component is negligible (i.e., does not affect \NHI\ more than $\approx5\%$), it can be ignored. Otherwise, the corresponding metallicity is treated as a lower limit (for 7 systems). Moreover, we identify one system with no detectable \HI\ associated with the \CIV\ transitions (see Appendix~\ref{sec:Q1317_lowlim}). Including this case, a total of 8 systems in our sample yield metallicity lower limits (``Lowlims''). For the remaining 44 systems, the modeling provides well-constrained metallicities (i.e., ``Measured'').

\begin{figure}
    \centering
    \includegraphics[width=\linewidth]{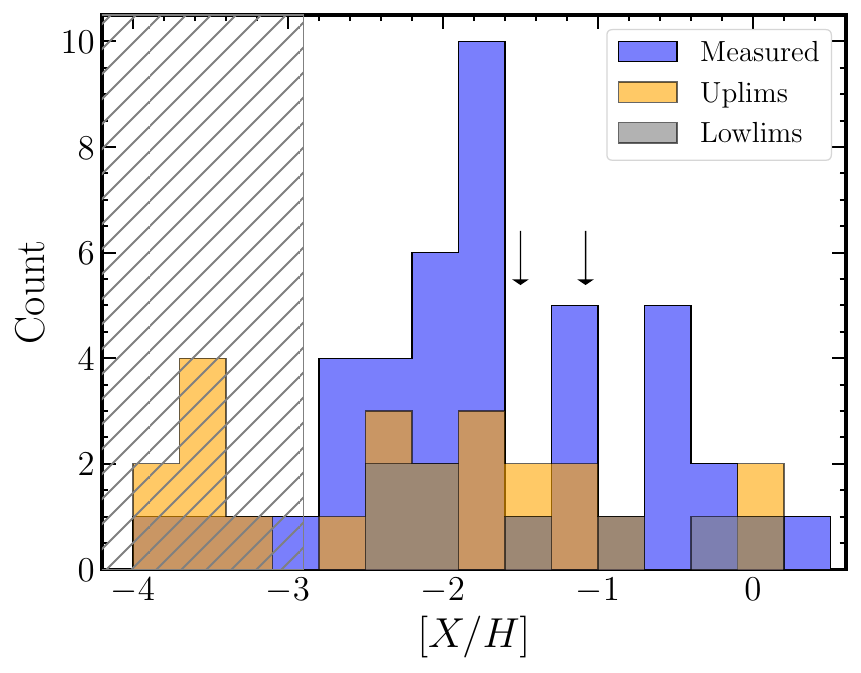}
    \caption{ Metallicity distribution of the absorbers associated with MUSEQuBES LAEs. Systems with well-constrained metallicities (``Measured'': purple), Upper limits (``Uplims'': yellow); and lower limits (``Lowlims'': gray), are shown separately. The vertical arrows indicate the metallicities of two DLAs, that were derived from \SiII\ abundance, {unlike the rest of the systems for which the Bayesian method was employed.}. The gray hatched region indicates typical IGM metallicities at these redshifts \citep{Schaye_2003}. Nearly $15\%$ of the absorbers lie in this region.} 
    \label{fig:met_hist}
\end{figure}

The resulting metallicity distribution is shown in Fig~\ref{fig:met_hist}. Systems with well-constrained metallicities, upper limits, and lower limits are indicated separately, with the two DLAs in our sample highlighted with arrows. Applying the Kaplan–Meier (KM) estimator to this censored dataset\footnote{8 systems with metallicity lower limits are excluded from this calculation.}, we derive a median metallicity of \met$=-1.6$, with $-2.5$ and $-0.4$ representing the $16^{\rm th}$ and $84^{\rm th}$ percentile values, respectively.

Out of the 37 systems for which \nh\ was estimated using Gaussian priors, 7 yield only metallicity upper limits and 8 provide lower limits, while the remaining 22 have direct measurements. The median metallicity of this subset is \met$=-2.2^{+0.5}_{-0.8}$; which is slightly lower than that of the full sample. The fact that these systems are, on average, more metal-poor, likely explains the absence of detectable transitions other than \CIV.

\citet{Schaye_2003} analyzed the statistical correlation between \CIV\ and \HI\ absorption using high-resolution QSO spectra over $1.5 < z < 4.5$, linking the absorption features to gas density and temperature via hydrodynamical simulations. Using hydrodynamical simulations and a UVB model, they translated the median \CIV/\HI\ pixel optical depth ratio into a median carbon abundance $[\rm C/H]$ as a function of density and redshift. Using their model\footnote{We adopt the best-fitting parameter values from their Table 2, specifically for the ``QG'' model UVB that includes both galaxy and quasar contributions.}, we estimate the median IGM metallicity of $[\rm C/H]\approx -3.7\pm0.8$, at $z=3$, for an overdensity of $\delta=1.4$, corresponding to $\log\, N({\rm HI})/ {\rm cm^{-2}}=14$, typical of IGM absorbers \citep{Schaye_2001}. The gray hatched region indicates this regime in the figure. A total of 11 systems ($\approx 15\%$) fall within this range. In addition, there are 16 metallicity upper limit estimates, which lie outside the shaded region, but the limits allow for the possibility that they are also consistent with the IGM metallicity. Including these brings the fraction up to 40\%.

\begin{figure}
    \centering
    \includegraphics[width=\linewidth]{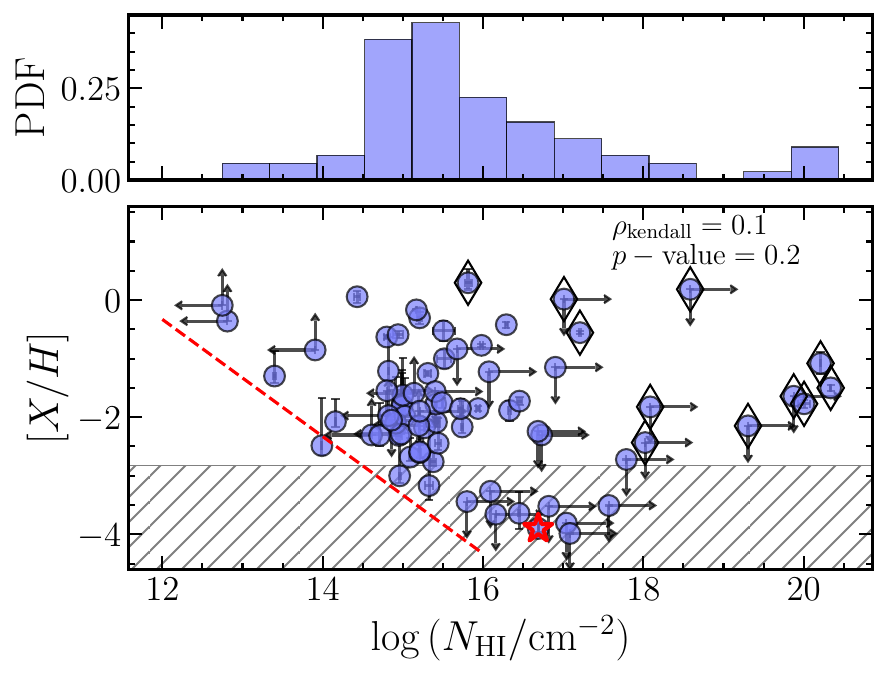}
    \caption{ The top panel shows the normalized \HI\ column density distribution of MUSEQuBES absorbers for which the metallicities could be constrained while the bottom panel shows their metallicities vs. \NHI. The diamond symbols in bottom panel mark systems with detected low-ion transitions; arrows indicate upper or lower limits on metallicity ($\downarrow/\uparrow$) and \NHI\ ($\leftarrow/\rightarrow$). The red dashed line marks the metallicity detection threshold set by the sensitivity of the quasar spectra (see text for details). No significant correlation is observed between the two parameters, as confirmed by the Kendall$-\tau$ test ($p-$value $= 0.2$). The gray hatched band indicates the typical IGM metallicity at $z \approx 3$. The extremely metal-poor absorber associated to the  nebula from \protect\cite{Banerjee2025_filament} is highlighted with a star symbol.
    } 
    \label{fig:nHI_met_scatter}
\end{figure}

The top panel of Fig.~\ref{fig:nHI_met_scatter} presents the \NHI\ distribution of the 75 absorbers whereas the bottom panel shows the metallicities as a function of \NHI. The results from a Kendall-$\tau$ correlation test using survival analysis reveals no statistically significant correlation between metallicity and \NHI. Absorbers with detected low-ion transitions are marked with diamonds. As previously noted, they arise from systematically higher-\NHI\ systems. 

The apparent absence of systems in the lower-left corner in Fig.~\ref{fig:nHI_met_scatter}, can be explained by the metal column density detection limits of the quasar spectra. In particular, the lowest metallicity we can determine is limited by the minimum metal column density that can be robustly detected by the quasar spectra. The median $3 \sigma$ limiting column density of \CIV\ in our spectra is $N$(\CIV) $\approx 10^{11.7}$~\sqcm\ \citep{Banerjee_2023}. For a given \NHI, this threshold can be used to estimate the lowest metallicity at which a system would still be detectable. We further quantified this using {\tt Cloudy} ionization models. At the redshifts of interest, and for a representative hydrogen number density of $\log\,(n_{\rm H}/{\rm cm^{-3}}) = -3.5$ (noting that the \CIV\ ion fraction peaks at this density for absorbers with \NHI$\lesssim 10^{17}$~\sqcm), we find that the ionization fraction ratio $f_{\CIV}/f_{\HI}\sim 10^{3.6}$, and is insensitive to both \NHI\ and metallicity. Using these values, one can calculate the metallicity detection threshold for the given spectral sensitivity using the following equation:
\begin{equation}
Z \;=\; \frac{N_{\mathrm{C\,IV}}/N_{\mathrm{H\,I}}}{(C/H)_{\odot}\,\left(\frac{f_{\mathrm{C\,IV}}}{f_{\mathrm{H\,I}}}\right)}
\end{equation}
Here, $(C/H)_{\odot}$ is the solar abundance of carbon from \cite{Asplund2009}. This sensitivity limit on metallicity is indicated by the red dashed line in the figure, which clearly explains the lack of data points in the lower-left corner of that plot.

We show the median IGM metallicity regime at $z=3$ (same as in Fig.~\ref{fig:met_hist}) by the gray hatched region. The 27 absorbers previously mentioned, which have metallicities consistent with that of the IGM, mostly arise in higher \NHI\ (\logN$\gtrsim16$) clouds. These high-\NHI\ clouds either exhibits metallicities consistent with the IGM or show higher metallicities, in which case they often display low-ion detections (diamond symbols). Note that, for the latter, metallicities are derived from the low-ion phase, whereas the low-metallicity systems are constrained using the high-ion phase. Notably, the IGM-like subsample includes the extremely metal-poor absorber associated with the giant \lya\ nebula discovered by \citet{Banerjee2025_filament}, highlighted with a star symbol.

\subsection{Variation of metallicity with galaxy environment}
\label{subsec:met_env}
\begin{figure}
    \centering
    \includegraphics[width=\linewidth]{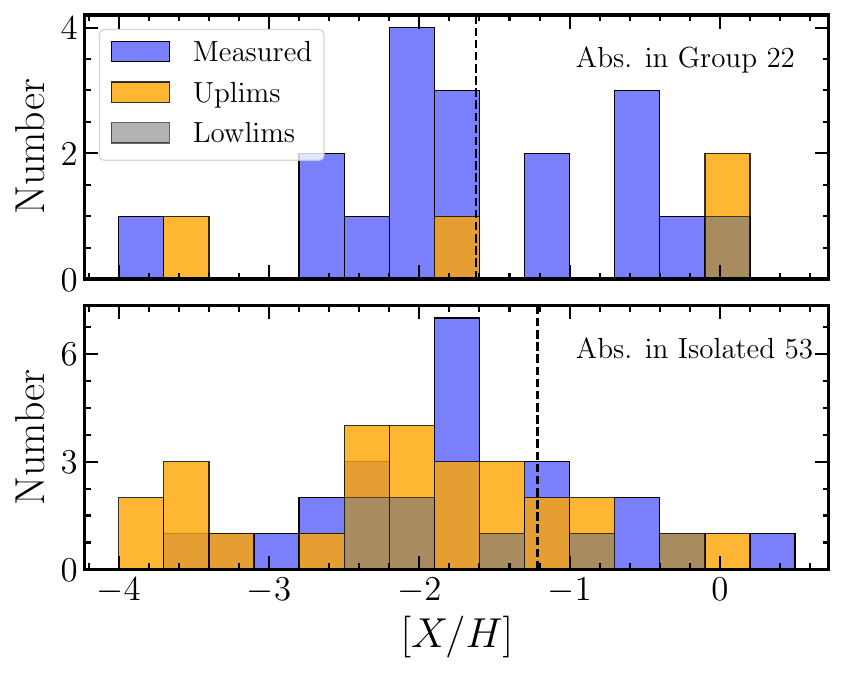}
    \caption{{\tt Top:} Metallicity distribution of the 22 absorbers associated with pairs/groups. The measured metallicities (purple), upper limits (yellow) and lower limits (gray) are shown separately. The black vertical dashed line marks the median metallicity from the Kaplan–Meier estimator, ignoring the lower limits. {\tt Bottom:} Same as the $\tt Top$ but for the 53 absorbers associated with isolated LAEs. No significant difference in the metallicity distributions is observed between these sub-samples.}
    \label{fig:met_env}
\end{figure}

Previous results from the MUSEQuBES survey have shown that neutral hydrogen and metals in group environments are enhanced compared to isolated galaxies \citep{Muzahid_2021, Banerjee_2023, Banerjee2025_HI}. This motivates us to explore whether the environment of LAEs influences the metallicity distribution of the surrounding gas.

To investigate this, we adopt a definition of ``group'' consistent with our earlier studies. Specifically, we sort the galaxies in each field by redshift and identify the LAEs that have at least one companion within $\pm500$~\kms\ in line-of-sight (LOS) velocity (and within the MUSE FoV $\lesssim 300$~pkpc) as ``pairs/groups''. LAEs without any such companions are classified as ``isolated''. Based on this criterion, we find that among the 75 absorbers for which metallicities could be constrained, 22 of them are associated with 29 group/paired LAEs and 53 are associated with 30 isolated LAEs.

\begin{figure*}
    \centering
    \hbox{
    \includegraphics[width=0.34\linewidth]{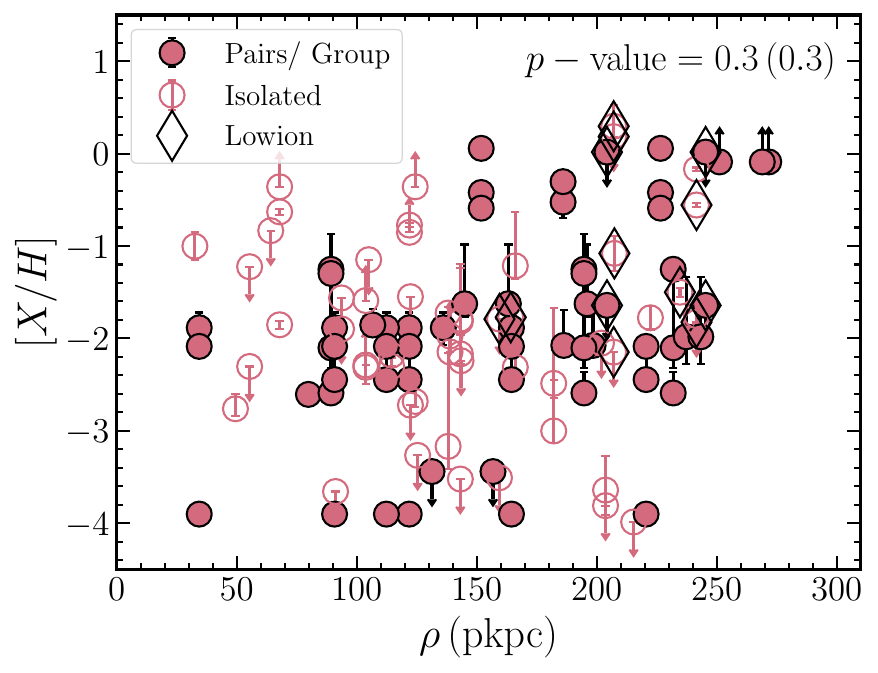}
    \includegraphics[width=0.3\linewidth]{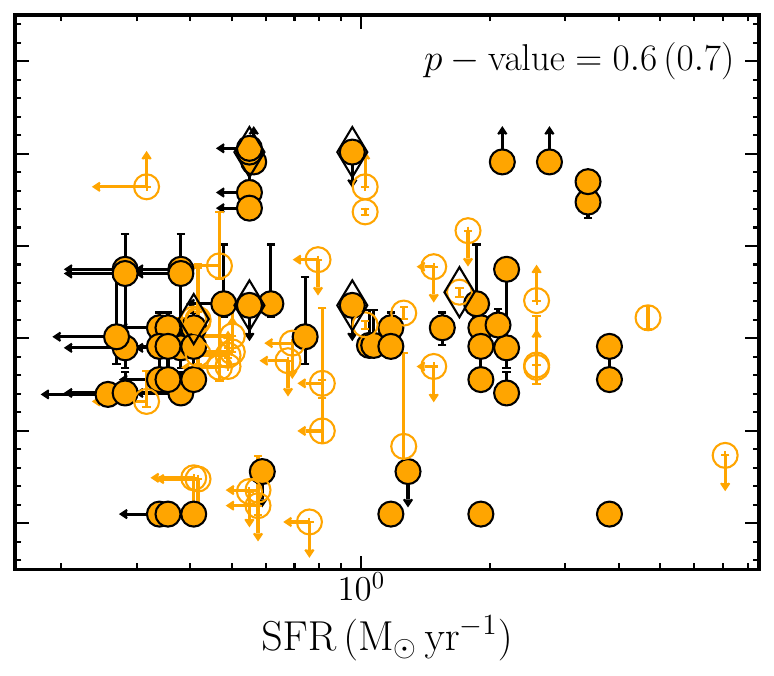}
    \includegraphics[width=0.3\linewidth]{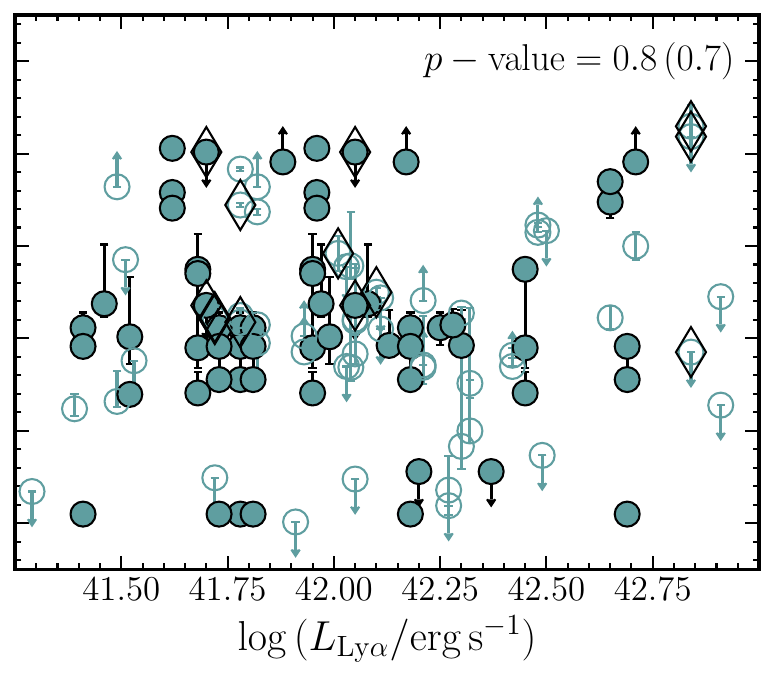}
    }
    \caption{Absorber metallicity plotted against galaxy properties: impact parameter (left), SFR (middle), and \lya\ luminosity (right). Filled circles denote absorbers associated with LAE pairs/groups, while open circles indicate those linked to isolated LAEs. Diamonds highlight systems with detected low-ion transitions, and arrows indicate upper or lower limits. Each panel lists the $p$-values from Kendall–$\tau$ correlation tests that account for censored data; values inside parentheses correspond to isolated LAEs, while those outside correspond to pair/group LAEs. No significant correlations are found between absorber metallicity and any of these parameters.}
    \label{fig:met_params}
\end{figure*}

Figure~\ref{fig:met_env} shows the metallicity distributions of these absorbers segregated by their environment. The color coding is the same as in Fig.~\ref{fig:met_hist}. The median metallicities, estimated using the Kaplan–Meier estimator (vertical dashed lines), are \met$=-1.6^{+1.6}_{-0.9}$ for groups and \met$=-1.2^{+1.0}_{-1.1}$ for isolated systems. A log-rank test, which accounts for censored data, yields a $p$-value of 0.7, indicating that the two distributions are statistically consistent. Notably, about 5\% of absorbers in pairs/groups and 11\% of those associated with isolated LAEs have metallicities consistent with the IGM. When metallicity upper limits are included, these fractions increase to 18\% and 43\%, respectively.

\subsection{Variation of metallicity with galaxy properties} 
\label{subsec:met}


In this section, we investigate whether the metallicity of absorbers associated with LAEs correlates with the intrinsic properties of the galaxies. We note that no significant redshift evolution of metallicity is observed in our sample, with a Kendall–$\tau$ test yielding a $p$-value of 0.6. This lack of evolution is likely due to the relatively small redshift interval probed by our sample, $\Delta z/(1+z) \approx 0.2$. Therefore, we do not account for redshift effects in the subsequent analysis. 


{Applying our fiducial galaxy–absorber association method, we find that 19 absorbers are associated with more than one LAE, while the remaining 56 are associated with individual LAEs.} The left panel of Fig.~\ref{fig:met_params} shows the metallicity of the systems as a function of the impact parameter of the associated LAEs. The absorbers associated to the isolated (group) LAEs are shown by empty (filled) circles. The plot shows the $p$-values from Kendall–$\tau$ correlation tests that includes the censored data: values inside parentheses correspond to isolated LAEs, while those outside correspond to pair/group LAEs. These values indicate that gas metallicity and impact parameter are not correlated.

The middle panel of Fig.~\ref{fig:met_params} presents metallicity as a function of the SFRs of the corresponding LAEs. We have excluded the LAEs that are blended with foreground sources as their SFR values might be affected by the unreliable continuum estimates. The majority of the remaining sample exhibits SFRs below $1\, \rm M_{\odot}\, yr^{-1}$. Once again, the Kendall-$\tau$ test, which accounts for dual censoring (i.e., censoring in both the x- and y-variables), reveals no significant correlation between metallicity and SFR. However, a difference is seen in the fraction of metallicity upper limits: approximately 15\% of systems with SFR~$>1\, \rm M_{\odot}\, yr^{-1}$ are upper limits, compared to $\approx$27\% for systems with SFR~$<1\, \rm M_{\odot}\, yr^{-1}$. However, the median metallicities of these two subsamples (accounting for the censored data) are very similar ($\approx -1.9$).

The right panel of Fig.~\ref{fig:met_params} shows metallicity against the luminosity of \lya\ emission (\lum). As for the other two panels, we do not observe a statistically significant correlation between metallicity and \lum.

\section{Discussions}
\label{sec:Discussions}

\subsection{Metallicity measurements}
\label{sec:met_discuss}

Metals are produced in stars and dispersed into the surrounding medium, and their distribution across cosmic time provides a record of the cumulative impact of star formation on the CGM and IGM. Tracing the metallicity of gas in different environments and epochs is therefore essential for understanding the buildup of chemical enrichment in the universe. In this work, we focused on gas around LAEs at $z=3$–4, prior to the era when cosmic star formation peaks ($z\approx2$).

Previous absorber-centric studies have established that the high-$z$ IGM is predominantly metal-poor. For example, \citet{kodiaq_z} found metallicities of \met$=-2.62$ for strong \lya\ forest systems (SLFSs; \NHI$<10^{16.2}$~\sqcm) and \met$=-2.23$ for pLLSs and LLSs (\NHI$>10^{16.2}$~\sqcm), both with nearly 1 dex scatter. Similarly, \citet{Fumagalli2016} reported \met$\approx-2.3\pm 0.8$ for LLSs and \met$\approx-1.9\pm 0.8$ sub-DLAs at these redshifts\footnote{We quote these values from \citet{kodiaq_z}, who recalculated these quantities using the same UVB as in this work.}. These values are consistent, within the uncertainties, with those of absorbers located within $\pm500$~\kms\ of the MUSEQuBES LAEs, which have a median metallicity of \met$=-1.6^{+1.2}_{-0.9}$. Our metallicity estimates are, also consistent with those reported by the MAGG survey \citep{Lofthouse_2023}, which also used absorbers associated to MUSE-selected galaxies. Their analysis found metallicities as low as \met$\approx -2.4^{+1.0}_{-0.6}$ for absorbers associated with at least one galaxy. It is important to note that their sample is targeted toward higher \NHI\ systems such as LLSs.

\begin{figure}
    \centering
    \includegraphics[width=\linewidth]{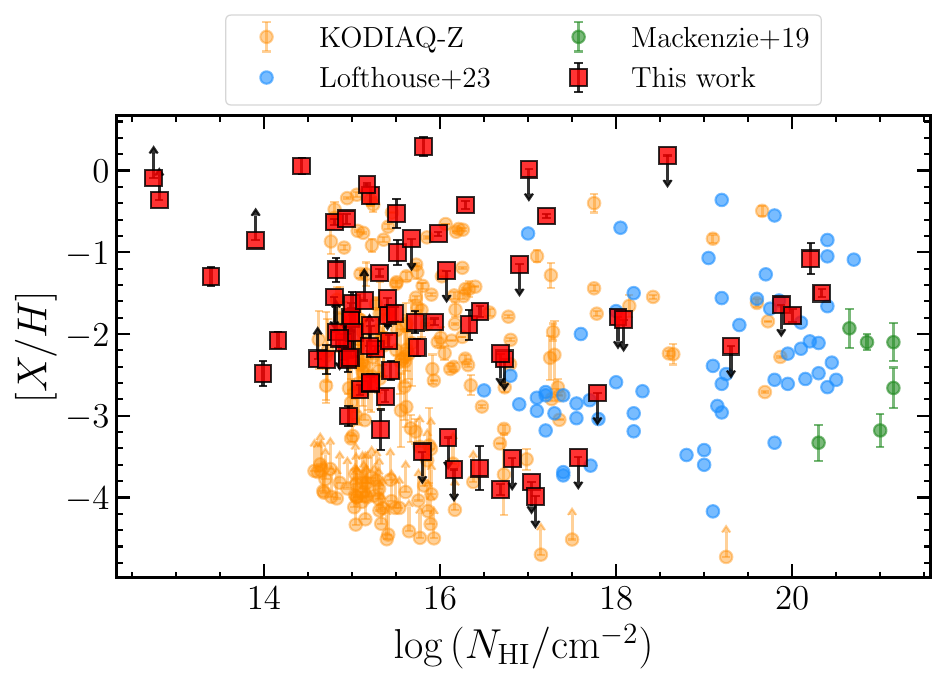}
    \caption{Metallicity as a function of \NHI\ for $z\approx2-$4 absorbers. Data from this work (red squares) are compared with previous surveys by \citet{kodiaq_z, Mackenzie_2019, Lofthouse2020}. Our measurements are broadly consistent with earlier results. }
    \label{fig:compare_met}
\end{figure}

In comparison, about 70\% of the absorbers in our sample are SLFSs \citep[i.e., similar to ][]{kodiaq_z}, while 14\% have \NHI$>10^{17.2}$~\sqcm, including 2 DLAs (both falling in the higher end of metallicity distribution, \met$\gtrsim-1.5$). The studies based on hydrodynamical simulations \citep[e.g.,][]{Fumagalli2011, van_de_voort_2012} predict that a large fraction of LLSs at $z\gtrsim3$ trace cold-mode accretion of relatively pristine gas \citep[to know more about cold-mode accretion, see, ][]{Keres_2005}. In contrast, absorbers selected through DLAs at these redshifts generally yield higher metallicities \citep[e.g.,][]{Mackenzie_2019}, with four out of six systems having \met$\gtrsim -1.5$, likely because they probe denser regions in closer proximity to galaxies. A comparison of absorber metallicities from the literature is compiled in Fig.~\ref{fig:compare_met}, where our metallicities are found to be broadly consistent with the earlier studies.

We identify a significant population of metal-poor absorbers in our sample. Specifically, 11 out of 75 systems ($\approx15^{+6}_{-3} \%$, $68^{th}$ percentile confidence from Wilson score) show IGM-like metallicities i.e., \met$\lesssim -2.8$ \citep[at $z=3$;][]{Schaye_2003}. When upper limits are included, this fraction increases to $40^{+5}_{-6}\%$. Notably, about 60\% of these systems fall within the pLLS/LLS regime, consistent with the picture that many LLSs at high redshift trace relatively pristine accreting gas. 

\begin{figure}
    \centering
    \includegraphics[width=\linewidth]{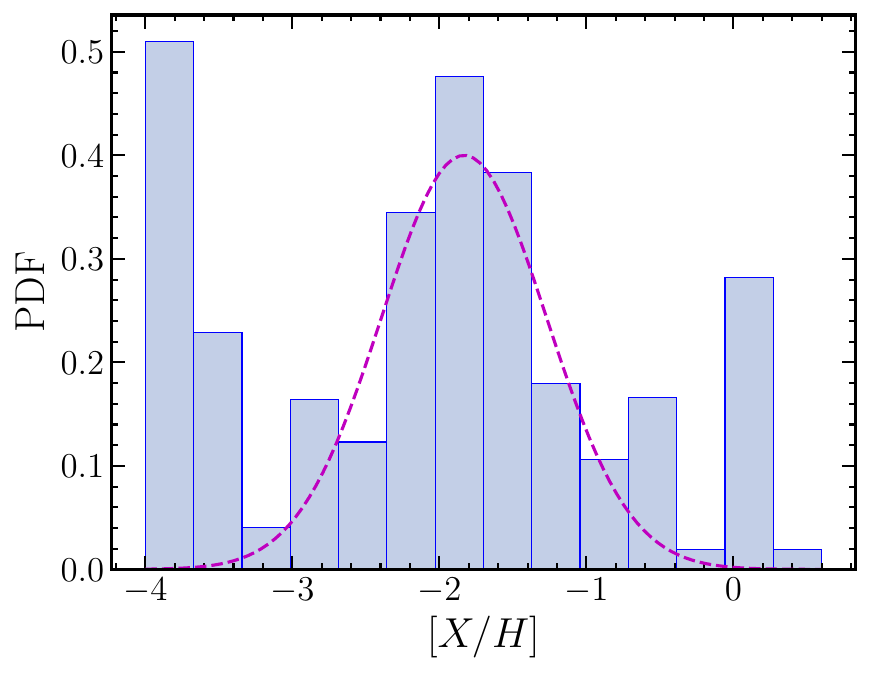}
    \caption{The metallicity posterior PDFs for all absorbers with $\log (N_{\rm HI}/ \rm cm^{-2}) > 16.5$. A prominent peak is evident at \met$\approx -1.8\pm0.6$, which is well captured by a Gaussian fit (magenta dashed line), suggesting a portion of the metallicity distribution is well-represented by a lognormal distribution. In contrast, a second peak at much lower metallicity (\met$\approx -3.8\pm0.2$) emerges, which is consistent with the IGM metallicity at these redshifts. }
    \label{fig:met_posterior_dist}
\end{figure}

To investigate this further, we focused on absorbers with relatively high \NHI, considering all systems with $\log (N_{\rm HI}/ \rm cm^{-2}) \gtrsim 16.5$ (corresponding to an overdensity of $\sim 100$ at $z=3$; \citealt{Schaye_2001}), resulting in a subsample of 19 absorbers. Their normalized metallicity distribution, constructed by combining the individual posterior PDFs, is shown in Figure~\ref{fig:met_posterior_dist}\footnote{The combined distribution includes posterior PDFs from both constrained metallicity measurements and cases with only upper or lower limits.}. We modeled this distribution using a Gaussian Mixture Model (GMM)\footnote{Implemented with {\sc scikit-learn} in Python.}, which fits the data with a chosen number of Gaussian components and compares the models using the Bayesian Information Criterion (BIC). The best-fit GMM reveals three distinct peaks: a prominent one at \met$\approx -1.8\pm 0.6$ (magenta dashed line), another at much lower metallicity (\met$\approx -3.8\pm 0.2$), and an additional peak near solar metallicity (\met$\approx 0 \pm 0.3$). The lowest-metallicity peak, consistent with typical IGM values at these redshifts, likely traces pristine gas in the cosmic filaments in which the LAEs are embedded \citep[see, e.g.,][]{Banerjee2025_filament}. The intermediate-metallicity peak presumably represents mixed gas in the CGM, where metal-rich outflows interact with pristine inflows from the IGM. The solar-metallicity peak, though tentative, would, if confirmed, point to fresh episodes of enriched outflows reaching the CGM before having time to mix with the ambient medium \citep[e.g., ][]{Scahye2007,Tripp2011, Muzahid2015, Rosenwasser2018}. Together, these results suggest that the CGM at high redshift is shaped by the interplay of inflows, mixing, and feedback, with distinct metallicity modes offering a window into the baryon cycle that governs galaxy evolution.


Finally, even though, in previous MUSEQuBES studies \citep{Muzahid_2021, Banerjee_2023, Banerjee2025_HI} both gas and metals were seen to be influenced by galaxy environment \citep[see also,][]{Galbiati_2023, Lofthouse_2023}, in the current work, metallicity does not appear to correlate with the number of nearby LAEs. \citet{Lofthouse_2023} likewise found that systems with multiple associated LAEs had mean and median metallicities comparable to those with only one LAE (\met$ \approx -2.4$). This apparent lack of an environmental trend in metallicity may reflect inefficient mixing processes, leading to a clumpy and chemically inhomogeneous gas distribution around LAEs. 


%
\begin{figure}
    \centering
    \includegraphics[width=\linewidth]{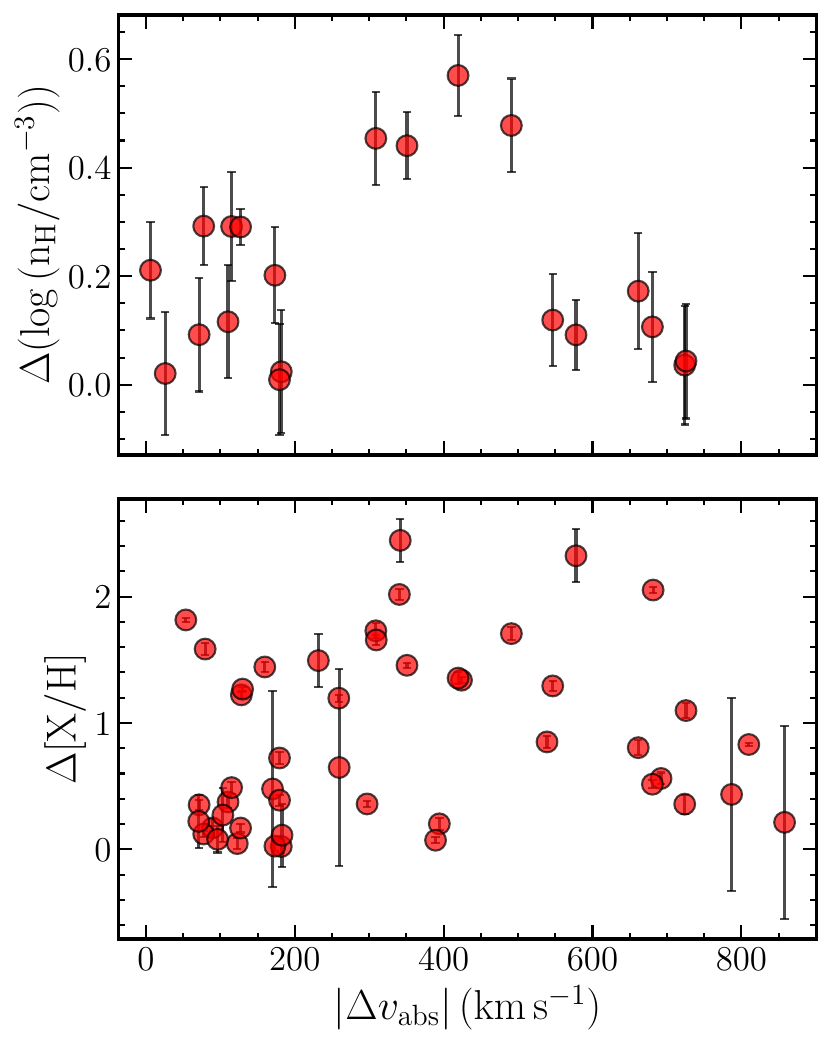}
    \caption{Differences in hydrogen number density (\nh; top) and metallicity (\met; bottom) between absorption systems separated by very small velocity separations.  The x-axis shows their LOS velocity separation. The y-axis error bars are derived through error propagation. Despite these small velocity separations, metallicity differences often exceed 1~dex. The \nh\ panel includes only those pairs with reliably constrained number densities using flat priors.}
    \label{fig:variation}
\end{figure}

\subsection{Variation of \met\ and \nh\ within small velocity separations} 
\label{subsec:variation}

Among the MUSEQuBES absorbers modeled in this work, several are separated by small LOS velocity differences (few $100$s~\kms), and in many cases, these absorbers are associated with the same LAE(s). Such configurations provide a valuable opportunity to probe inhomogeneities in density and metallicity within the CGM along the LOS. To explore this, we constructed all possible absorber pairs within individual quasar sightlines and measured their velocity separations. Since our modeling is restricted to absorbers within $\pm500$~\kms\ of MUSEQuBES LAEs, applying a small velocity cut effectively selects pairs linked to the same isolated LAE or to members of the same pair/group of LAEs.

Figure~\ref{fig:variation} shows the resulting variations in hydrogen number density (\nh; top panel) and metallicity (\met; bottom panel) between absorber pairs with velocity separations $|\Delta v| < 500$~\kms. In the top panel, we include only pairs where both absorbers have reliably constrained densities (i.e., modeled with flat priors), leading to fewer points than in the metallicity panel. In the bottom panel, metallicity upper and lower limits are excluded for clarity. Nonetheless, \nh\ are seen to vary upto $\approx 0.5$~dex over these modest velocity separations. For metallicity, differences often exceed 1~dex. Such small-scale fluctuations are consistent with findings from previous studies \citep[e.g.,][]{Prochter2010, kodiaq_z}, and point towards the multiphase nature of CGM. However, note that, we have ignored the small velocity separations between \HI\ and metal components during modeling.

\subsection{Ionization parameters} 

For our sample, we find a median hydrogen number density of $\log\, (n_{\rm H}/\rm cm^{-3}) = -2.7 \pm 0.5$, which corresponds to an overdensity of $\approx200$, i.e., a galaxy-halo like overdensity. Further, this density corresponds to an ionization parameter of $\log\, U \approx -2.1 \pm 0.5$ for the assumed HM05 UV background at $z = 3$–4. This is broadly consistent with earlier findings. For example, \cite{Fumagalli2016} reported that the densities of LLSs at $z \approx 2.5$–3.5 span the range $\log\, (n_{\rm H}/\rm cm^{-3}) = -3.5$ to $-2$, with ionization parameters\footnote{ Their models are constrained from the typical ions observable for LLSs at these redshifts such as, \CII, \SiII, \AlII, \CIV, \SiIV\ \citep[see, ][]{Prochaska_2015}.} $-3 < \log\, U < -2$. Similarly, \cite{kodiaq_z} found an average hydrogen density of $\log\, (n_{\rm H}/\rm cm^{-3}) = -2.69 \pm 0.54$ for absorbers with $\log\, (N_{\rm HI}/\rm cm^{-2}) = 16.4$, closely matching our median values.

It is important to note that derived gas densities depend significantly on the assumed ionizing background in photoionization models. However, previous studies, including \cite{kodiaq_z}, have shown that different UV background models (e.g., \citet[][HM12]{Haardt-Madau2012}, \citet[][KS18]{Khaire2019}) yield broadly consistent results with HM05 at these redshifts (see their Figure 7). Thus, we expect the choice of UVB model to have a relatively minor impact on our density estimates.

That said, incorporating additional ionizing sources, such as radiation from nearby galaxies or quasars, can influence the inferred densities. For instance, \cite{Fumagalli2016} showed that including a local ionizing contribution can increase the inferred density by about $0.33 \pm 0.49$ dex. However, because our sample predominantly probes regions at projected distances greater than 50 pkpc from galaxies, it is unlikely that local radiation sources significantly impact our results \citep{Rahmati_2013}. Moreover, \cite{Fumagalli2016} also demonstrated that metallicity estimates remain largely unaffected by such additional ionizing contributions.

\section{Conclusion} 
\label{sec:Conclussion}

In this work, we investigated the physical properties of absorbers located within a few hundred pkpc (median impact parameter $\approx 160$~pkpc) of LAEs at $z \approx 3.3$, as part of the MUSEQuBES survey. To this end, we constructed a comprehensive grid of photoionization models using {\tt Cloudy}, adopting the HM05 UV background as the ionizing source. The models span a wide range of parameters, including (i) neutral hydrogen column density (\NHI), (ii) redshift ($z$), (iii) metallicity (\met), and (iv) total hydrogen number density (\nh). Using a Bayesian framework, we derived the posterior PDFs of these physical parameters.

Our findings can be summarized as follows: 

\begin{itemize}
    \item  We modeled 75 absorbers associated with 59 MUSEQuBES LAEs through a custom-built Bayesian framework. Nearly 70\% of these absorbers have $\log\, (N_{\rm HI}/\rm cm^{-2}) < 16.2$, while $\approx$15\% exhibit higher column densities with $\log N(\HI)/{\rm cm^{-2}}\gtrsim17.2$, including two damped \lya\ absorbers (DLAs). 

    \item The photoionization equilibrium models are primarily constrained using \HI, \CIV, and \SiIV\ column densities. A smaller subset of absorbers (10 out of 75) also exhibit low-ionization species such as \SiII, \FeII, \AlII, and \AlIII. These low ions do not always trace the same kinematics as the high ions, and when present, they are used exclusively to constrain the models. 
    
    \item For systems where the associated \NHI\ represents a lower limit, the resulting metallicities are treated as upper limits. Conversely, for systems with no detectable \HI\ or where a single \HI\ component is shared between multiple metal systems, yet contributes significantly to their total \NHI, the modeled metallicities are considered as lower limits. Applying survival analysis to this censored dataset, we find a median metallicity of \met~$\approx -1.6^{+1.2}_{-0.9}$ (see Fig.~\ref{fig:met_hist}).

    \item For a subset of systems, we could determine hydrogen number densities using flat priors in the Bayesian analysis. The median value is $\log\, (n_{\rm H}/\mathrm{cm^{-3}}) = -2.7 \pm 0.7$, corresponding to an ionization parameter of $\log\, U \approx -2.1 \pm 0.7$ for the HM05 UV background at $z = 3$–4.

    \item 11 out of 75 systems ($\approx15^{+6}_{-3} \%$, 68$^{th}$ percentile confidence from Wilson score) show IGM-like metallicities (\met$\lesssim -2.8$ at $z=3$). When upper limits are included, this fraction increases to $\approx40^{+5}_{-6} \%$. Notably, about 60\% of these systems have \NHI\ consistent with being a pLLS/LLS.

    \item The combined metallicity PDF of a subset of absorbers with $\log\, N($\HI$)/\rm cm^{-2} \gtrsim 16.5$ (corresponding to overdensities $\gtrsim 100$ at $z\approx3$) reveals three distinct populations: (1) a very metal-poor population (\met$\lesssim -2.8$) consistent with pristine IGM gas in cosmic filaments, (2) an intermediate-metallicity population tracing mixed CGM gas, and (3) a tentative solar-metallicity population indicating recently enriched outflows that have not yet mixed with the ambient medium (see Fig.~\ref{fig:nHI_met_scatter} and Fig.~\ref{fig:met_posterior_dist}). 

    \item We find no statistically significant difference in absorber metallicity between group and isolated environments (see Fig.~\ref{fig:met_env}). About $5\%$ of absorbers in pairs/groups and $11\%$ of those associated to isolated LAEs have metallicities consistent with that of the IGM (excluding the upper limits).

    \item There is no significant trend between metallicity and galaxy properties such as impact parameter, SFR, or \lya\ luminosity. This is also true for \nh, except for a notable correlation between \nh\ and impact parameter, which appears to be driven primarily by systems with detected low-ion transitions, possibly reflecting the missing hosts due to observational limitations (see Fig.~\ref{fig:met_params} and \ref{fig:hden_params}).

    \item Absorbers separated by small velocity intervals ($\lesssim 500$~\kms) often exhibit $\approx$1 dex variation in metallicity and $\approx 0.5$~dex in \nh, pointing to an inhomogeneous metal distribution in the gaseous atmospheres of these LAEs (see Fig.~\ref{fig:variation}).
    
\end{itemize}

As a natural next step, we plan to perform a detailed cloud-by-cloud modeling of the observed absorbers, following the approach of \citet{sameer2021}. This will enable us to determine the metallicity, density, temperature, and line-of-sight thickness of individual components, moving beyond the average properties considered in this work.

\section*{ACKNOWLEDGEMENT}
We thank Raghunathan Srianand for useful suggestions. This work has used IUCAA HPC facilities. This paper uses the following software: NumPy \cite[]{Harris_2020}, SciPy \cite[]{Virtanen_2020}, Matplotlib \cite[]{hunter_2007}, and AstroPy \cite[]{Astropy_2013, Astropy_2018}. 

\section*{Data Availability}

The data underlying this article are available in ESO (http://archive.eso.org/cms.html) and Keck (https://www2.keck.hawaii.edu/koa/public/koa.php) public archives.



\bibliographystyle{mnras}
\bibliography{zbib} 



\onecolumn
\appendix

\section{An example of our method to segregate absorption components into systems:}

\begin{figure}
    \centering
    \includegraphics[width=0.6\linewidth]{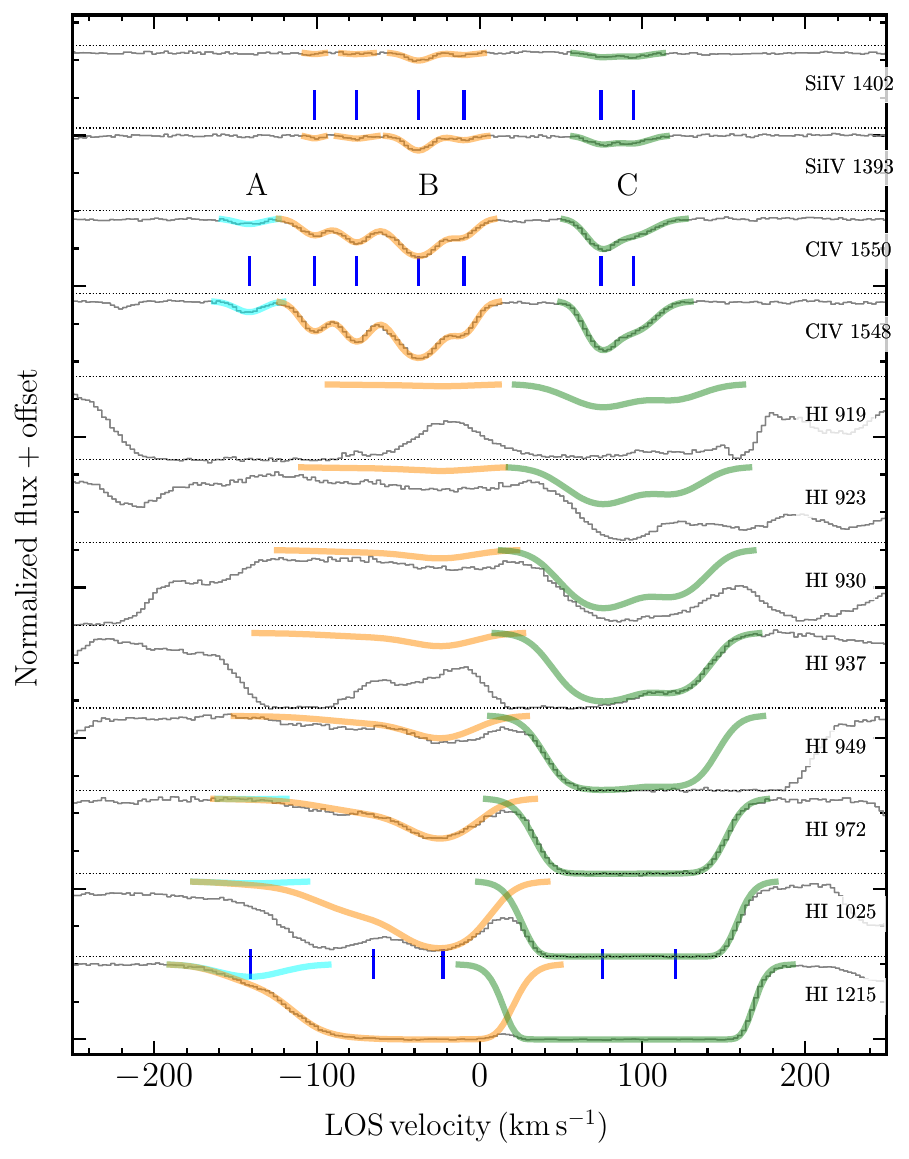}
    \caption{An example of our method to segregate one absorption feature along the quasar sightline Q0055$-$269, into multiple systems. Here it is split into three systems: systems A, B and C, in order of increasing redshift. In addition to \HI, all the systems exhibits clear detections of \CIV. Additionally, two of them (system B and C) have \SiIV\ detected. The small vertical ticks indicate the locations of the individual components. For all three of them, \NHI\ is well constrained with the help of higher order Lyman series lines.  
    }
    \label{fig:example_veloplot}
\end{figure}

Fig.~\ref{fig:example_veloplot} shows the spectra and its best fitting Voigt profile decomposition of all available Lyman series lines, together with the detected metal transitions, along the quasar sightline Q0055$-$269. The redshift of the galaxy with which the absorbers are associated corresponds to $\Delta v=0$~\kms. Following the method described in Section~\ref{sec:abs_measurment}, we segregate this into three ``systems'', closely separated in velocity space (hereafter, systems A, B and C, in order of increasing redshift). The combined Voigt profile of each of these systems is over-plotted on the quasar spectrum (in aqua for system A, orange for system B and green for system C). 

\section{Velocity plots for the DLAs from our survey}
\label{sec:Appen_DLAs}
\begin{figure}
    \hbox{\includegraphics[width=0.45\textwidth, trim=0 0 0 1cm,clip]{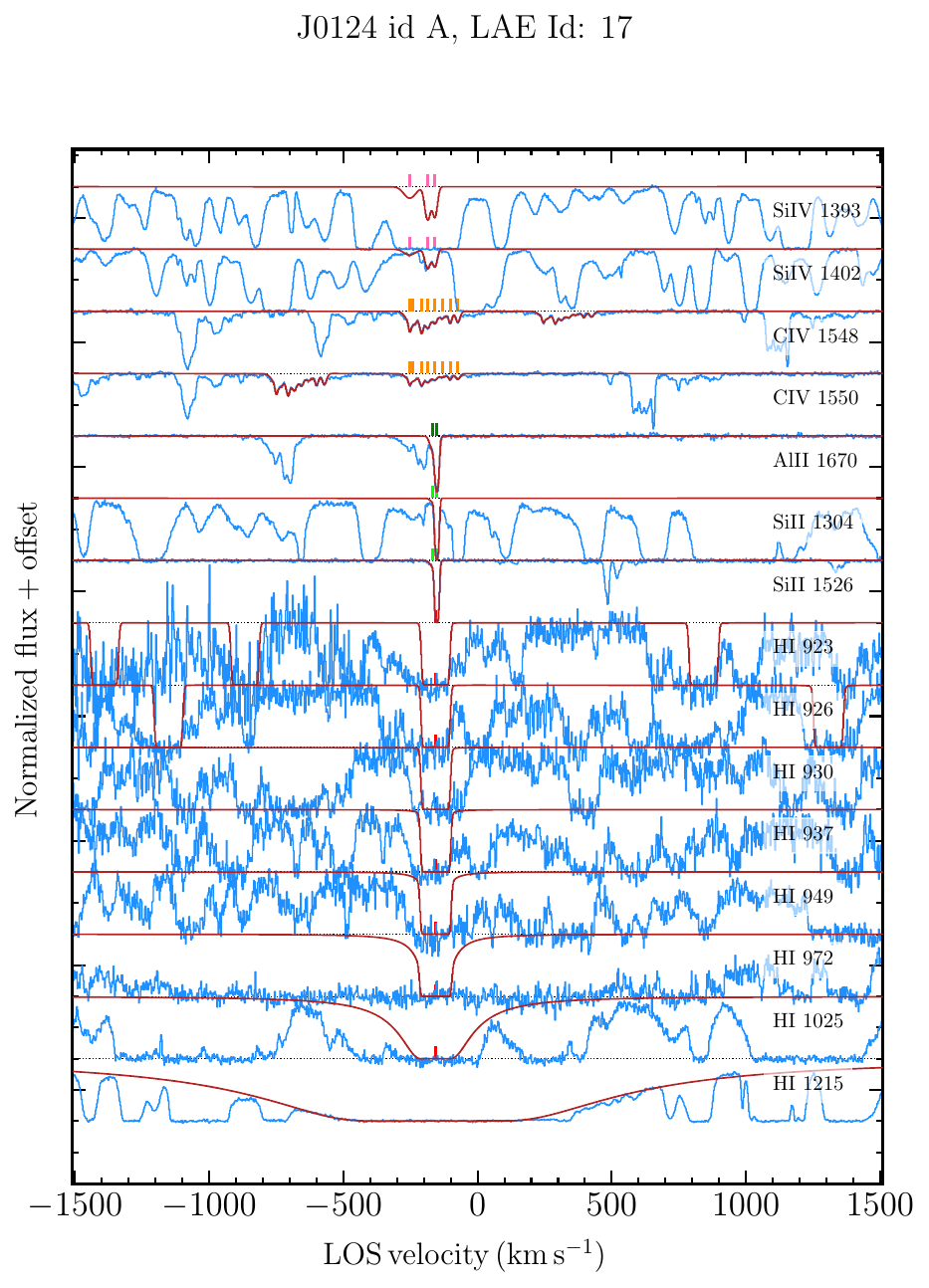} 
    \includegraphics[width=0.45\textwidth, trim=0 0 0 1cm,clip]{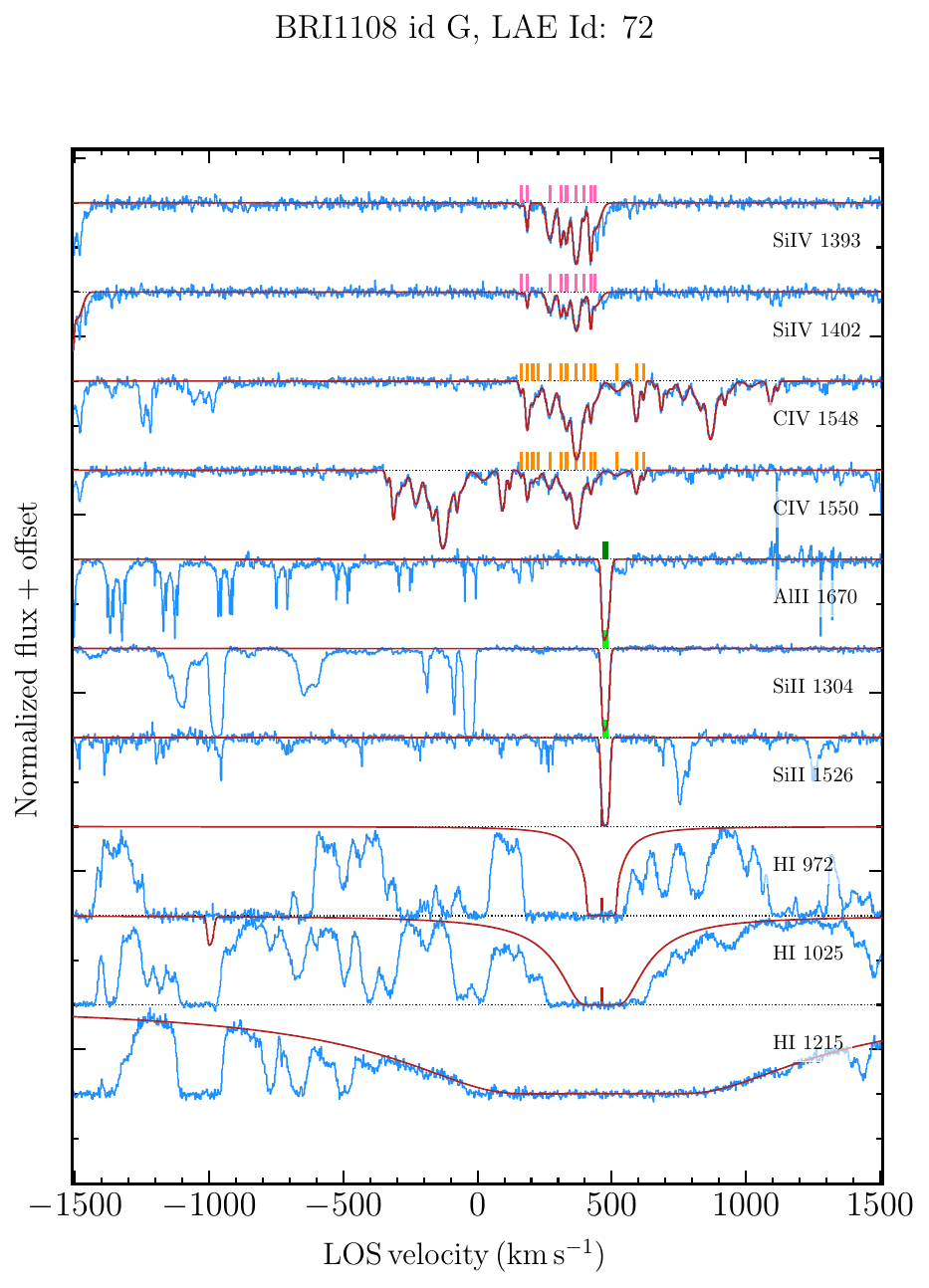}
    } 
    \caption{Velocity plots for the two DLAs from our survey. The observed spectra (blue) and best-fit Voigt profiles (maroon) are shown for selected Lyman series and metal transitions. Vertical ticks mark the positions of individual components. The $0$~\kms\ corresponds to the galaxy redshift.}
    \label{fig:DLA_vpfit}
\end{figure}

\begin{figure}
    \centering
    \includegraphics[width=0.6\linewidth]{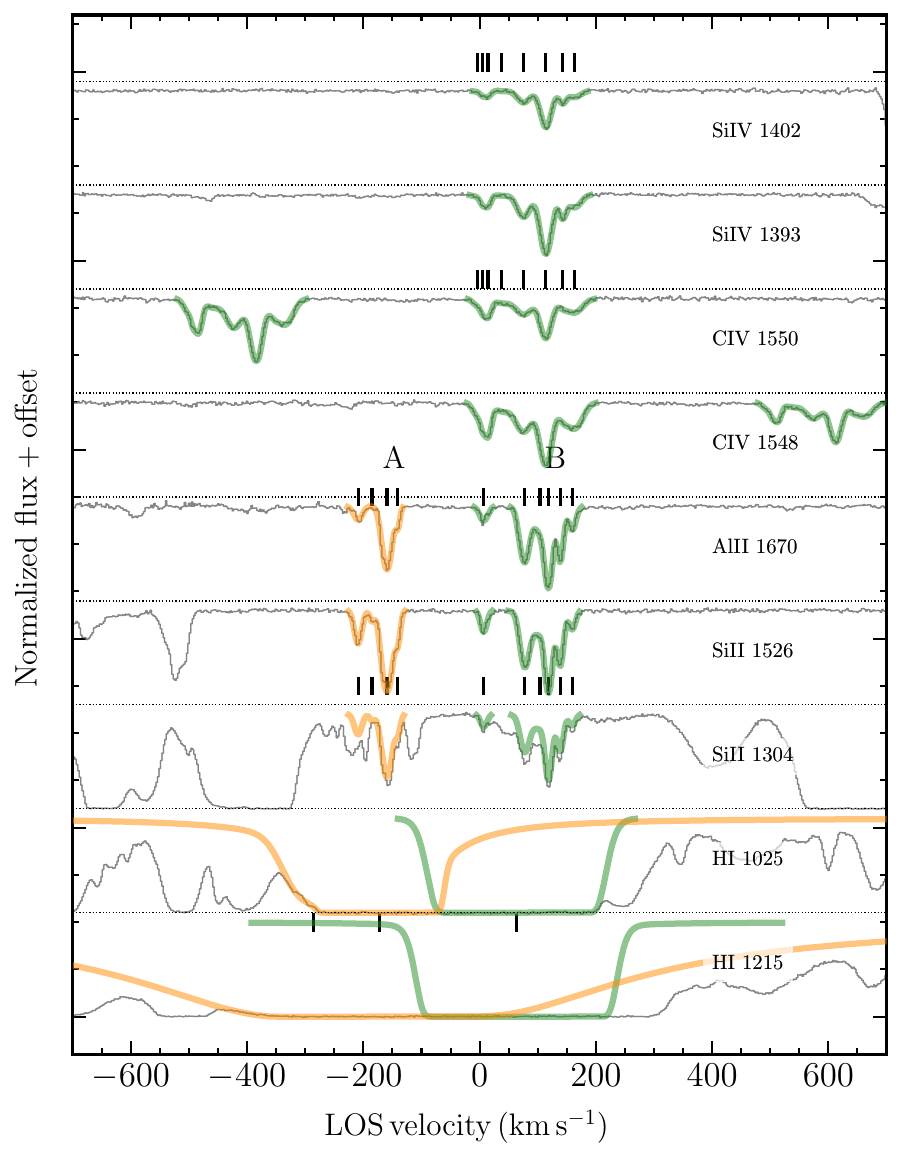}
    \caption{AN example of the spectra (in black) and best-fitting Voigt profile for two systems along the quasar sightline QB2000$-$330. The system on the left (marked A: orange) shows detections only of low ions, whereas the system on the right (marked B: green) exhibits detections of both low and high ions. The small vertical ticks indicate the locations of individual components. For system B, the low- and high-ion absorption follows a similar component structure in velocity space, in contrast to the two systems shown in Fig.~\ref{fig:DLA_vpfit}. }
    \label{fig:no_dif_low_n_high_ion_veloplot}
\end{figure}

Here, we show the velocity plots (Fig.~\ref{fig:DLA_vpfit}) for the Lyman series lines and the metal lines detected towards two DLAs; along 0124$+$0044 ({\tt left}) and BRI1108$-$07 ({\tt right}). The corresponding  best-fit HI column densities are \logN$=20.21$ and $20.33$, respectively. Note that, while low-ionization lines (e.g., \SiII\ and \AlII) are aligned with the \HI\ absorption, the higher-ionization transitions (e.g., \CIV\ and \SiIV) exhibit entirely different velocity and component structures. The metallicities of these systems are determined for the low-ionization gas phase. A full-blown multiphase model of these absorbers will be elsewhere.\\
While the examples above illustrate cases in which low- and high-ion species trace distinct velocity or component structures, this behavior is not ubiquitous. Fig.~\ref{fig:no_dif_low_n_high_ion_veloplot} shows two absorption systems in close velocity separation, labeled A (orange) and B (green). System B exhibits detections of both low and high-ions, which follow similar component structures, in contrast to the DLAs discussed earlier. System A shows detections only of low ions. The corresponding \NHI\ values are $\log N(\HI)=19.9$ for system A and $\log N(\HI)=17.6$ for system B. 
%


\section{Example of our modeling method:}
\label{sec:example_model}
In this section, we illustrate our modeling method with the systems represented in Fig.~\ref{fig:example_veloplot}. The example shows three systems: A, B and C. closely separated in velocity space. The total column density of each of these systems is used to constrain their respective model. 

Note that the \HI\ component shown in aqua is actually shared by both systems A and B. However, B contains other components with \NHI\ order(s) of magnitude stronger than that, whereas this is the only \HI\ associated to system A. As a result, the total \NHI\ associated to system A will be considered as an upper limit, hence its metallicity will be a lower limit. For system B, however, metallicity will not be censored, as the shared \HI\ has negligible effect on its total \NHI. See section~\ref{sec:abs_measurment} for details regarding the grouping of metal absorption components into systems and the assignment of \HI\ components to these systems. Fig.~\ref{fig:example_model} presents the corner plot of the posterior PDFs of the model parameters for each of these systems, along with a comparison between the model-predicted and observed column densities of the different transitions associated to this system.

\begin{figure}
    \centering
    \hbox{
    \includegraphics[width=0.5\linewidth]{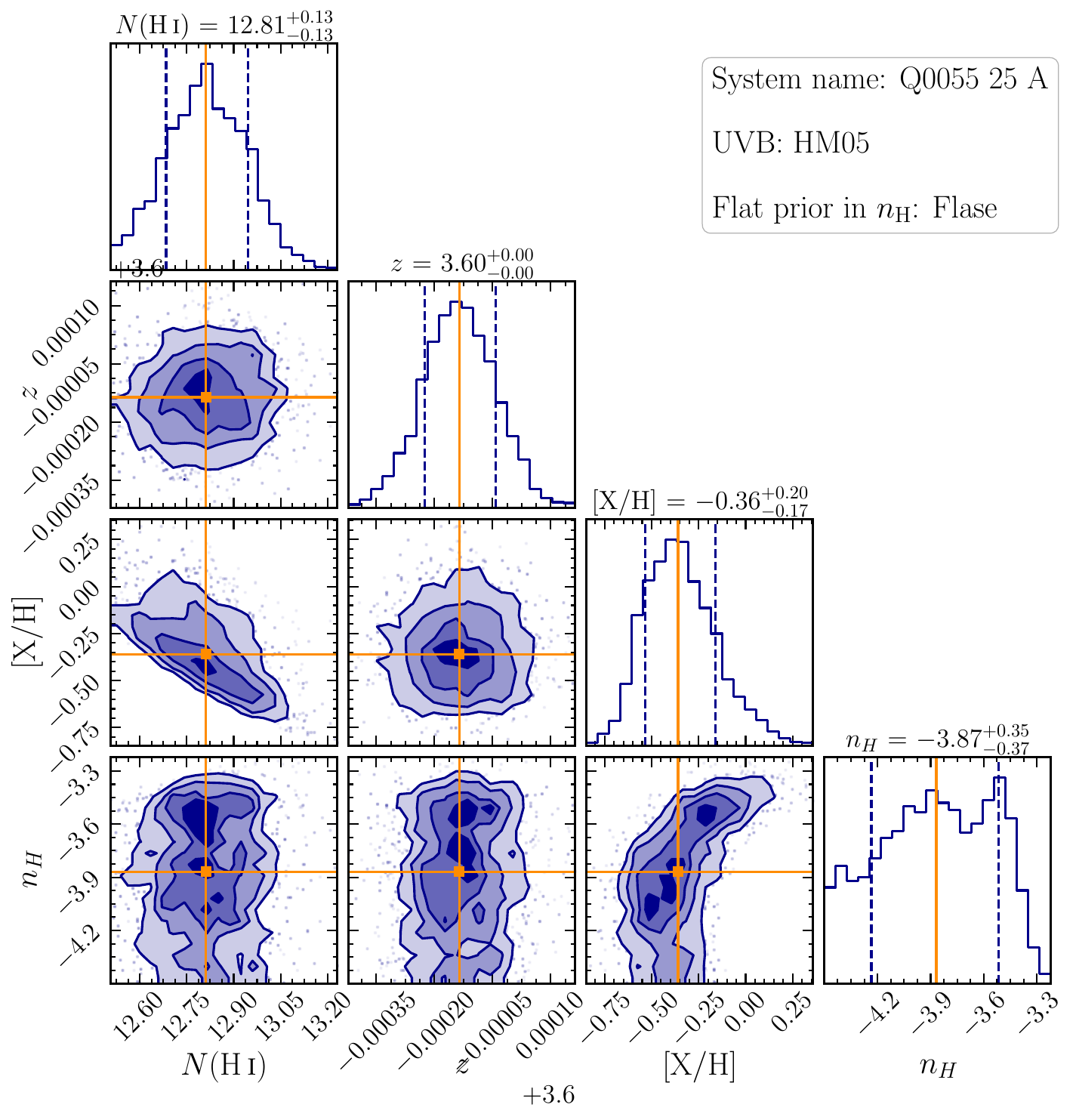}
    \includegraphics[width=0.5\linewidth]{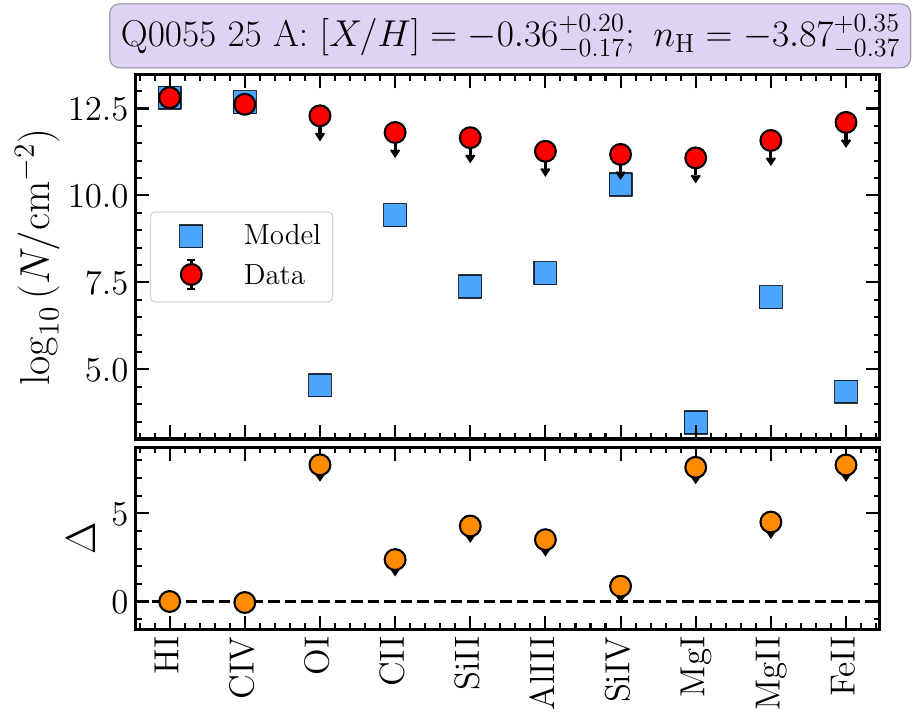}
    }
     \hbox{
    \includegraphics[width=0.5\linewidth]{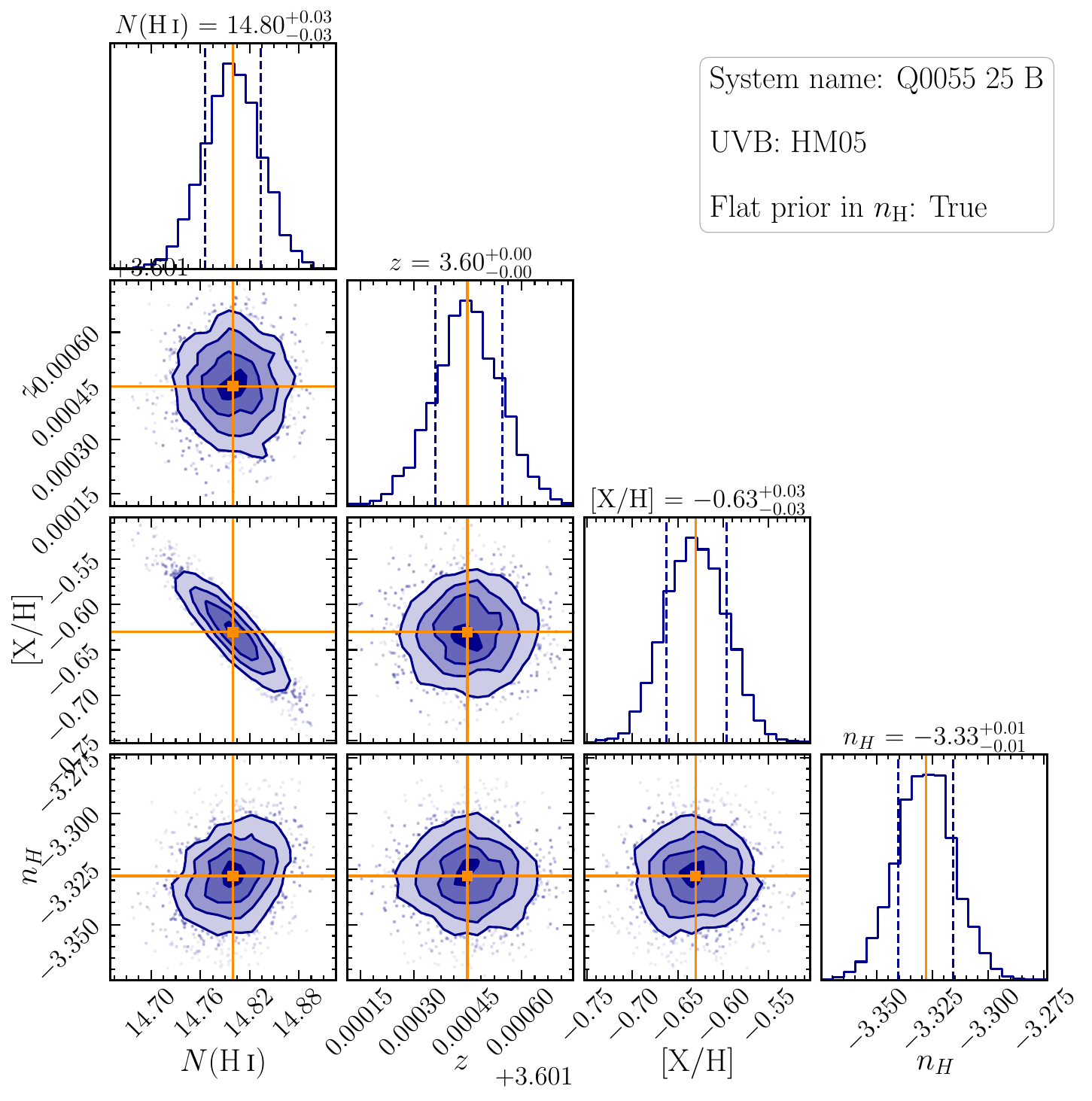}
    \includegraphics[width=0.5\linewidth]{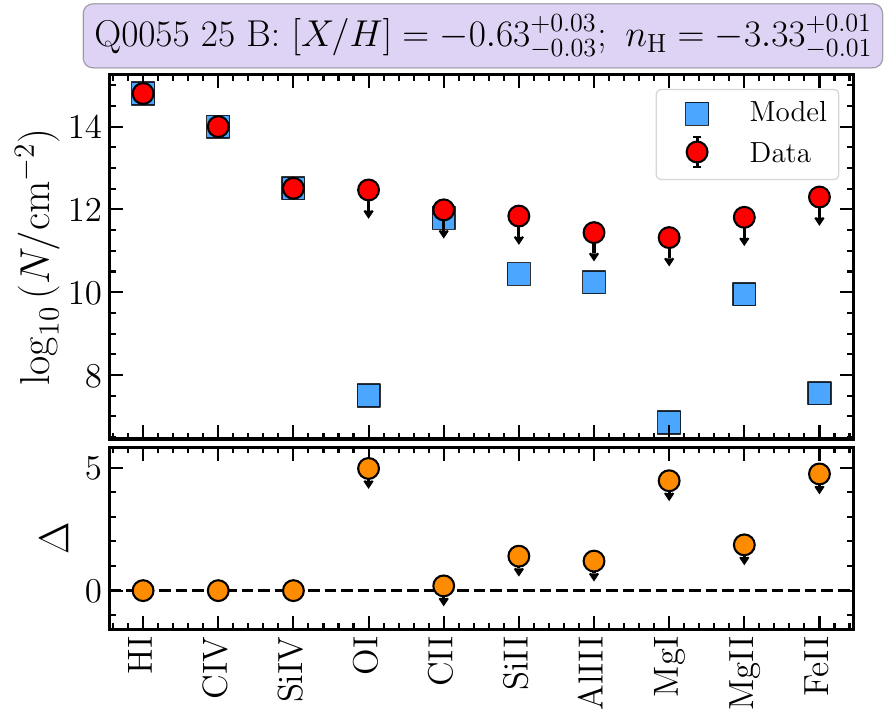}
    }
    \caption{This figure illustrates our Bayesian modeling approach for all three systems shown in Fig.~\ref{fig:example_veloplot}. In each panel, the left plot displays the corner plot of the posterior distributions for the four model parameters, while the right panel compares the measured (red points) and modeled (blue squares) column densities, along with the residuals. We could use a flat prior on \nh\ for systems B and C. However, since \CIV\ is the only metal transition detected in system A, we had to model it using a Gaussian prior on \nh. }
    \label{fig:example_model}
\end{figure}

\begin{figure}
    \centering
     \hbox{
    \includegraphics[width=0.5\linewidth]{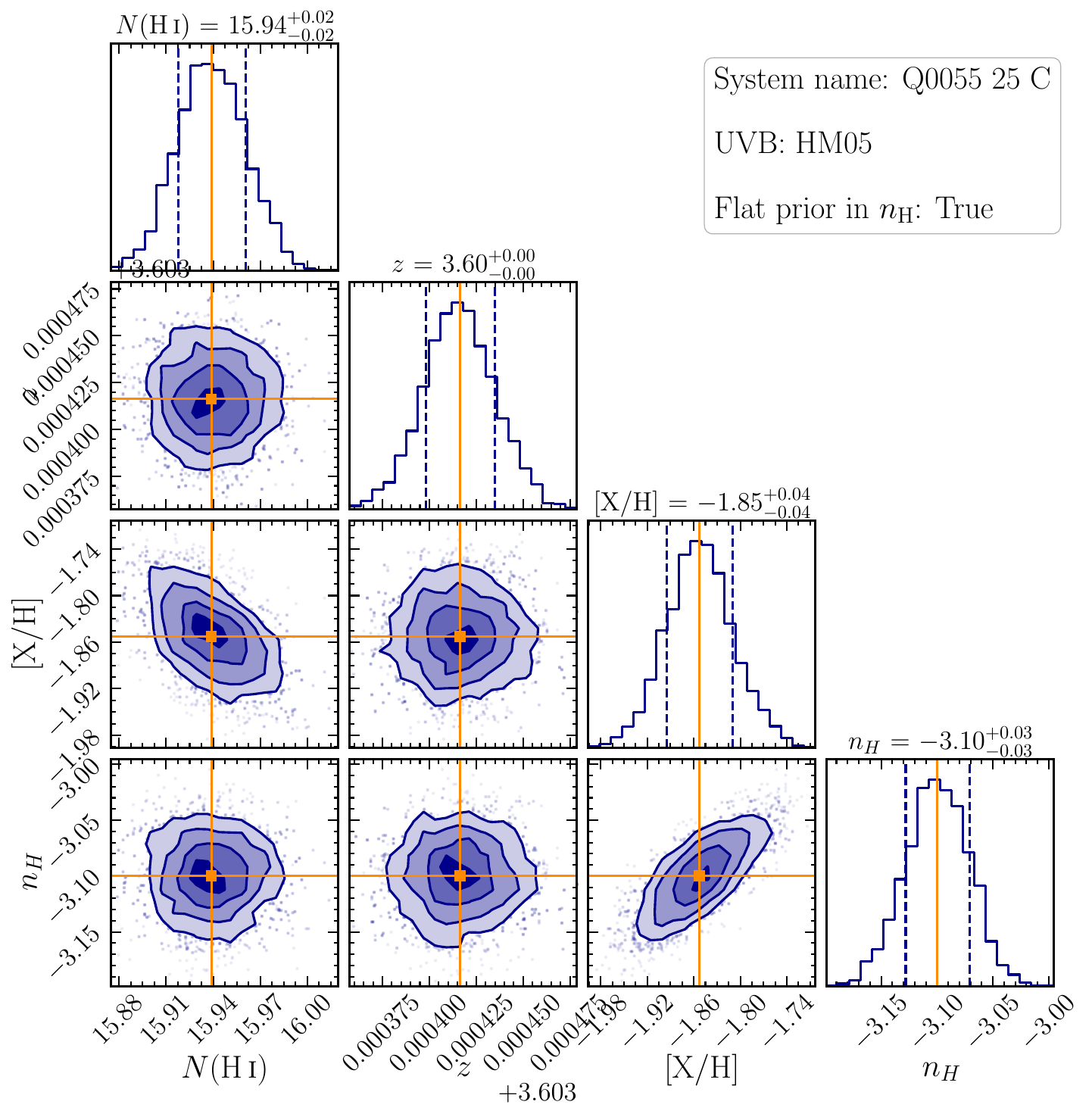}
    \includegraphics[width=0.5\linewidth]{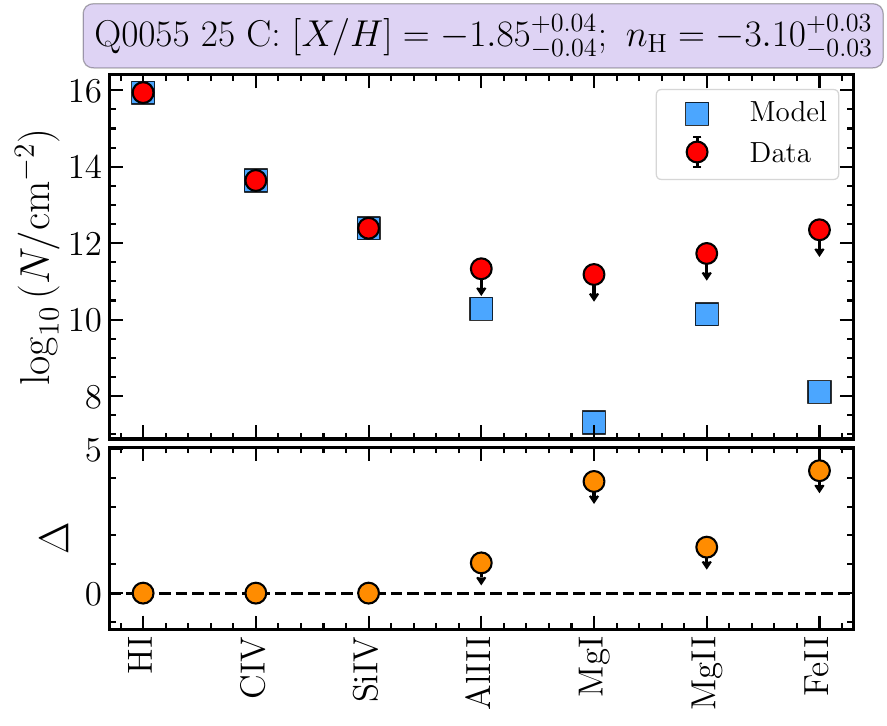}
    }
    \caption{Continuation of Fig.~\ref{fig:example_model} }
\end{figure}

\section{Metallicity and density as function of different galaxy properties}
\label{sec:appendix_galaxy_param}
\begin{figure*}
    \centering
    \hbox{
    \includegraphics[width=0.34\linewidth]{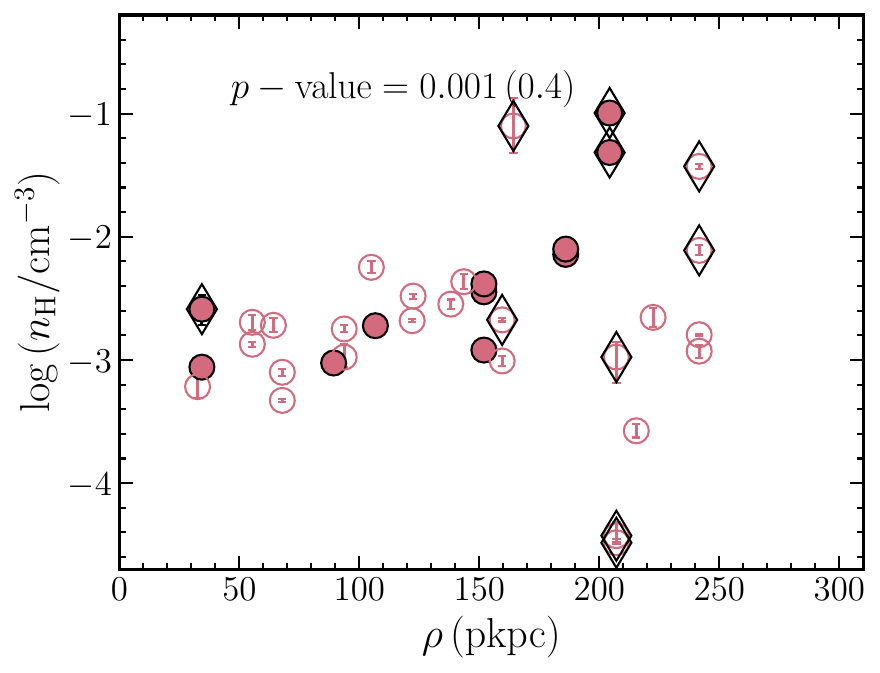}
    \includegraphics[width=0.3\linewidth]{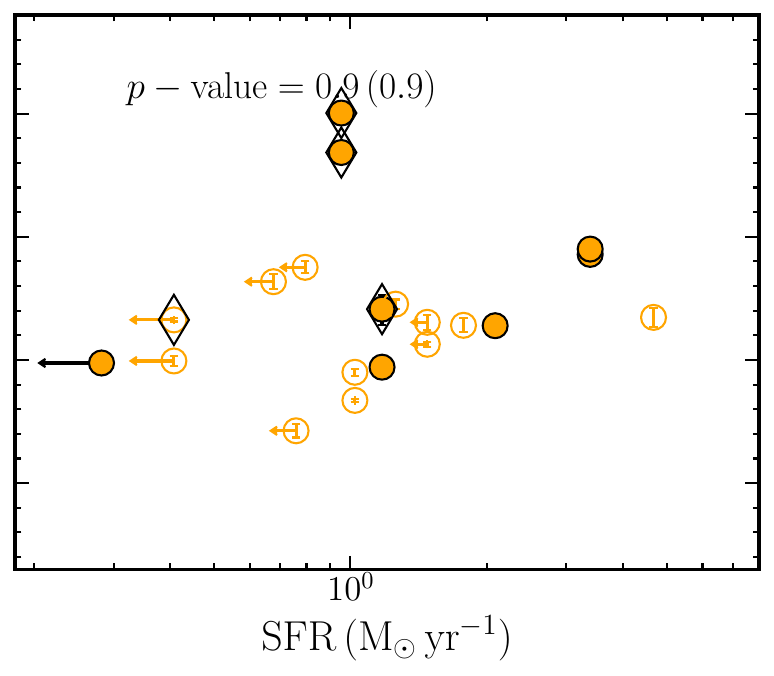}
    \includegraphics[width=0.3\linewidth]{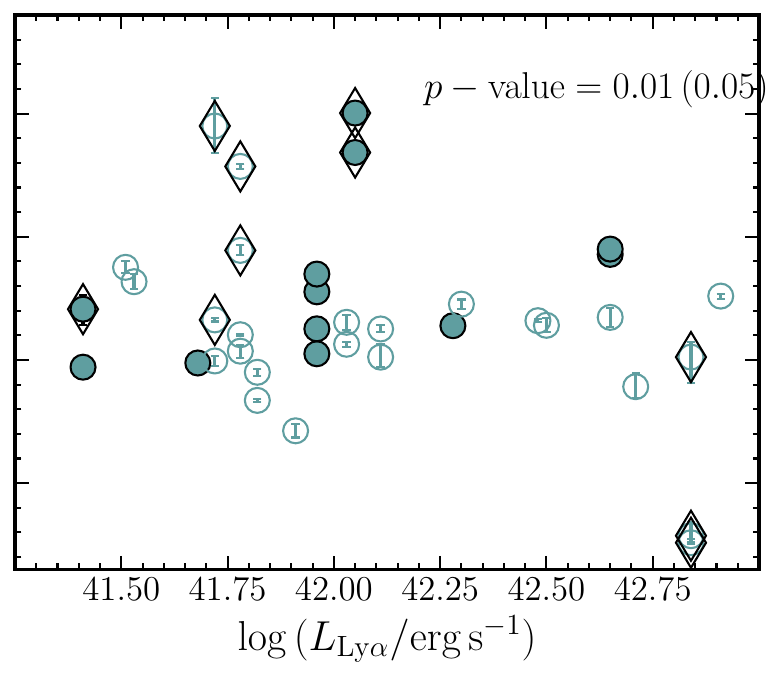}
    }
    \hbox{
    \includegraphics[width=0.34\linewidth]{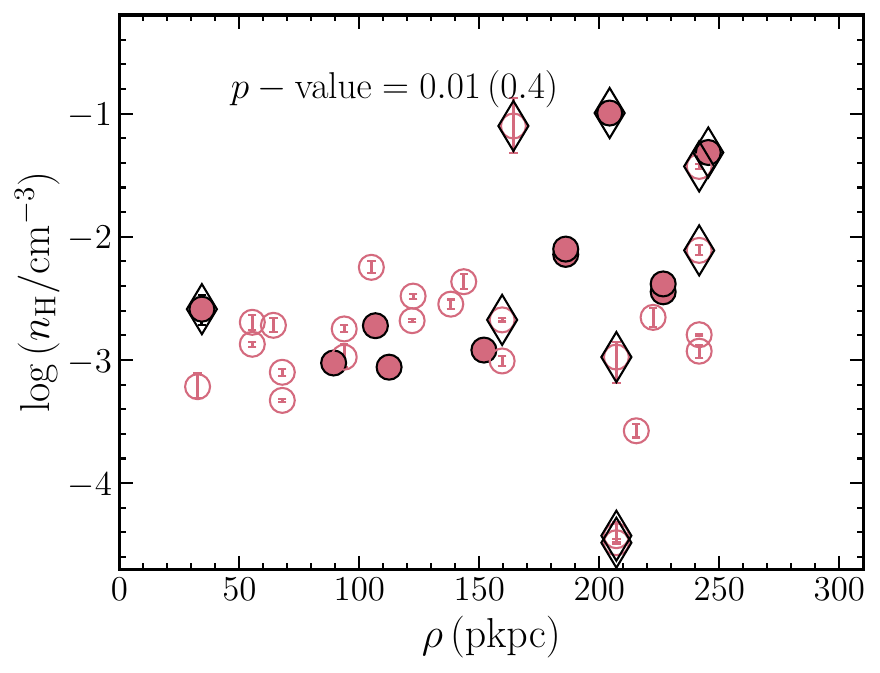}
    \includegraphics[width=0.3\linewidth]{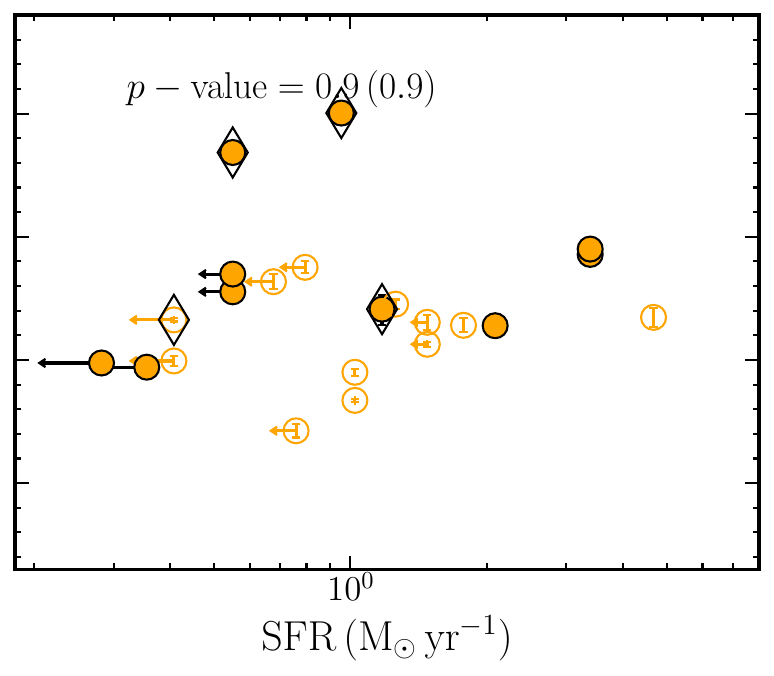}
    \includegraphics[width=0.3\linewidth]{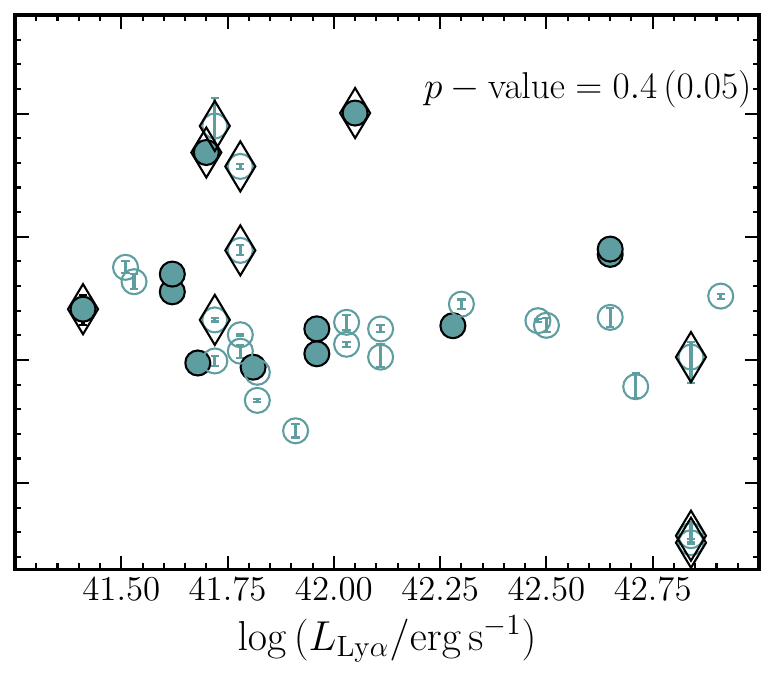}
    }
    
    \caption{Same as in Fig.~\ref{fig:hden_params}, but here when multiple LAEs are found within $\pm500$~\kms\ of the same absorber, we have assigned it to the LAE with lowest impact parameter ({\tt top} panel) and to the one with lowest 3D distance ({\tt bottom} panel). No significant trend is seen in any panel.}
    \label{fig:hden_params_appendix}
\end{figure*}

\begin{figure*}
    \centering
    \hbox{
    \includegraphics[width=0.34\linewidth]{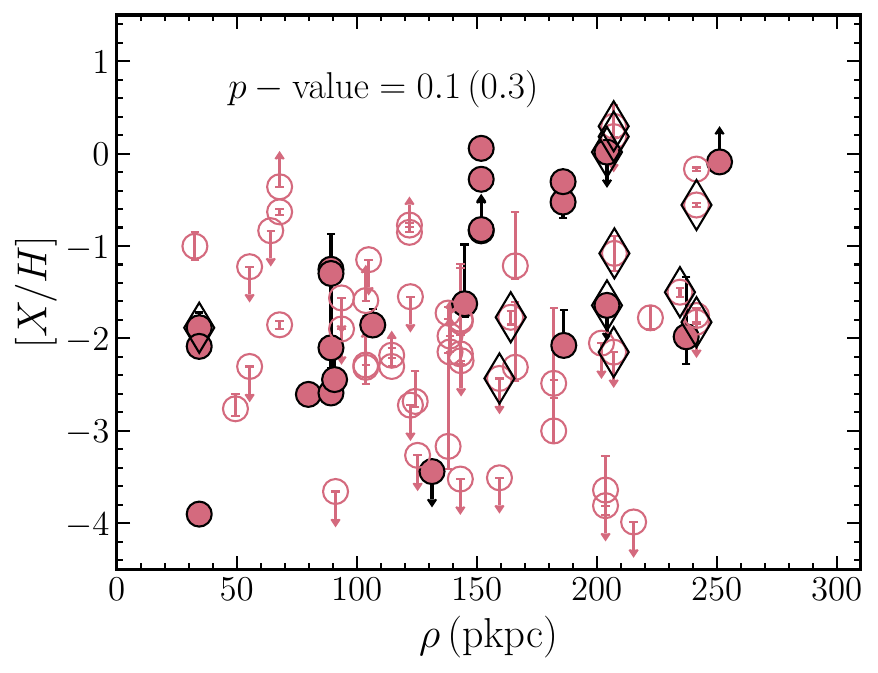}
    \includegraphics[width=0.3\linewidth]{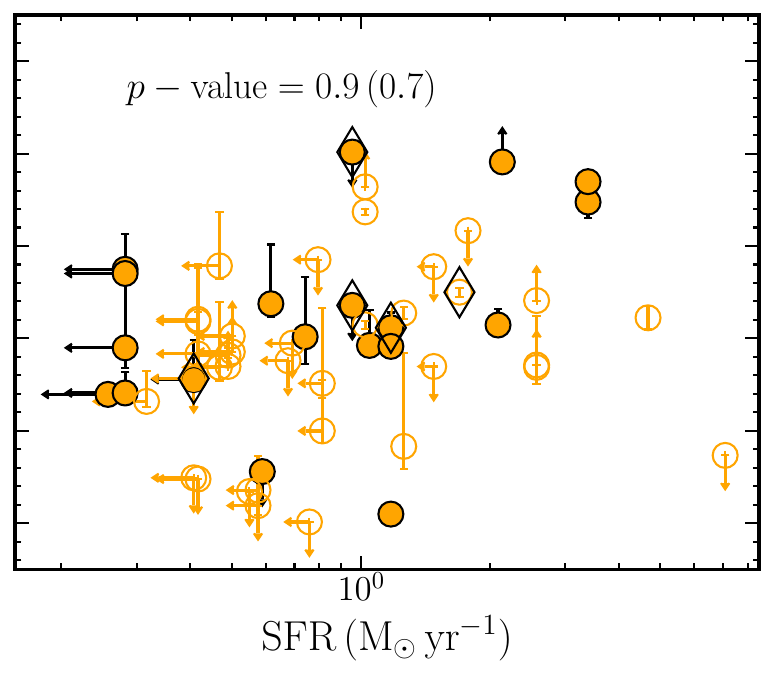}
    \includegraphics[width=0.3\linewidth]{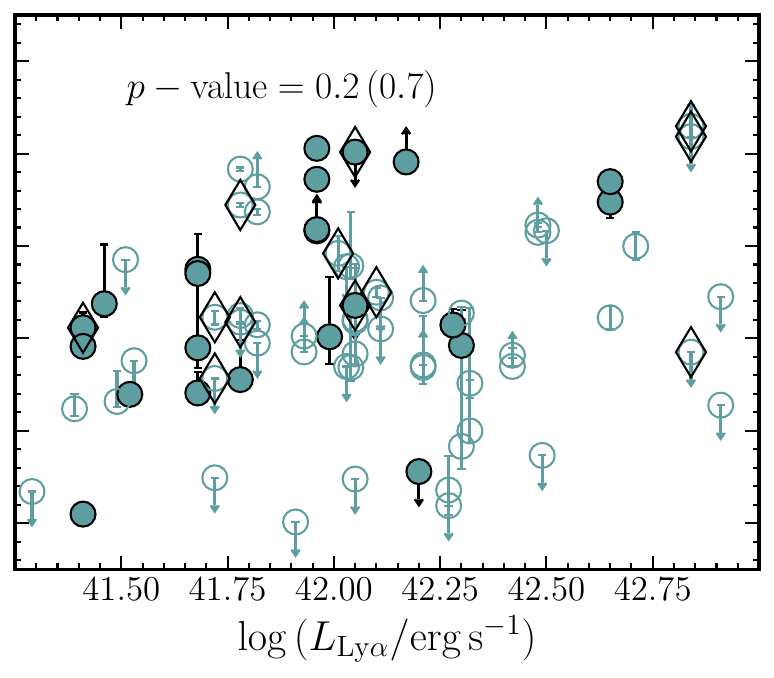}
    }
    \hbox{
    \includegraphics[width=0.34\linewidth]{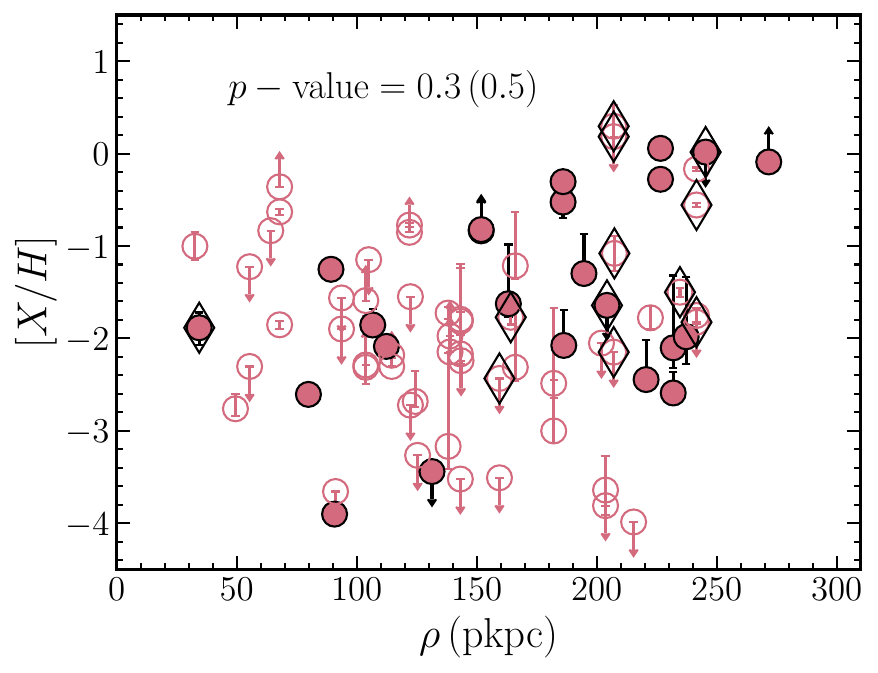}
    \includegraphics[width=0.3\linewidth]{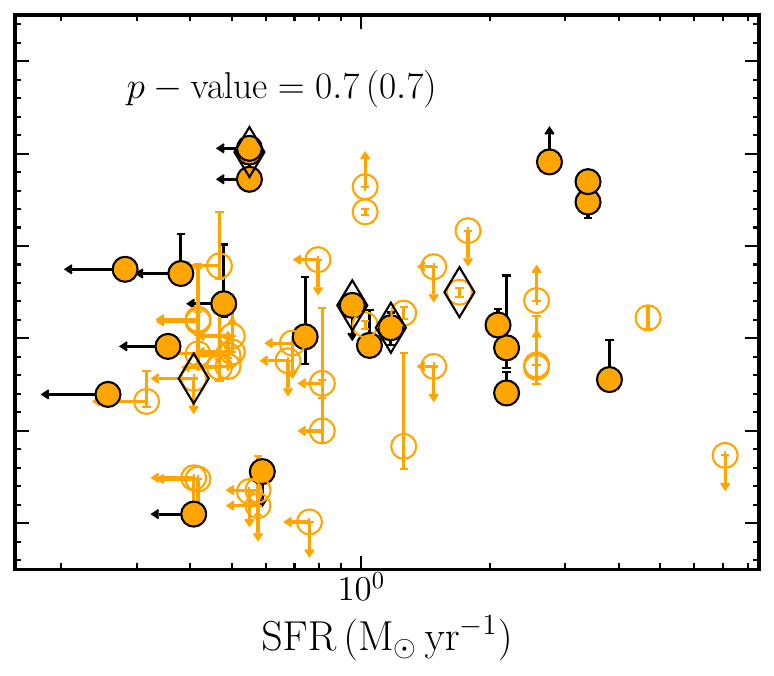}
    \includegraphics[width=0.3\linewidth]{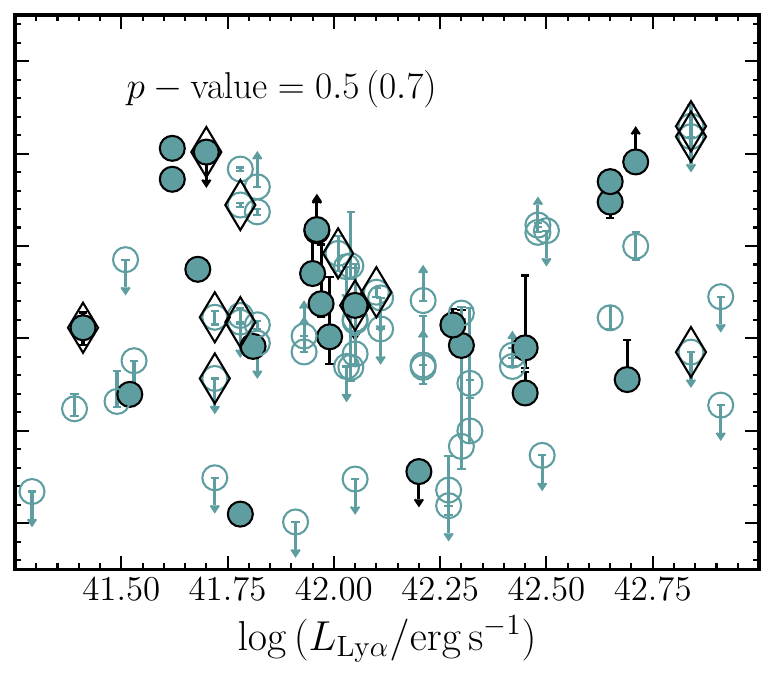}
    }
    \caption{Same as in Fig.~\ref{fig:met_params}, but here when multiple LAEs are found within $\pm500$~\kms\ of the same absorber, we have assigned it to the LAE with lowest impact parameter ({\tt top} panel) and to the one with lowest 3D distance ({\tt bottom} panel). No significant trend is seen in any panel.}
    \label{fig:met_params_appendix}
\end{figure*}

As mentioned in Section~\ref{subsec:hden}, in order to assign galaxies to absorbers, we have adopted three different approaches. The results from the fiducial approach where we assign an LAE to all the systems detected within $\pm500$~\kms\ of its redshift, has already been discussed in the main text. Here we are presenting the other two approaches i.e., in case of multiple associated LAEs within $\pm500$~\kms, we assign the system to the LAE at lowest impact parameter, and at lowest 3D distance. We do not find any new trend with these two approaches either, as can be seen from Fig.~\ref{fig:hden_params_appendix} and \ref{fig:met_params_appendix}.
%
\section{Robustness of the derived physical parameters}
\label{sec:appendix_toy_model}
In this paper, we are using the total column density of a system in order to constrain the modeling parameters i.e. metallicity and density, which is not ideal because different components might arise in different gas phases.

To assess the reliability of our modeling framework, we designed a set of experiments. First, we construct a hypothetical multi-component absorption system at 
$z \approx 3$, assigning each component a range of \NHI, metallicity, and density. Using the {\sc Cloudy} grids, we then predict the corresponding \CIV\ and \SiIV\ column densities for each component. We focus on these two ions since our models are primarily constrained by them. Next, we combined the column densities of the individual components and applied our Bayesian inference script to recover the corresponding density and metallicity of the system. This procedure allowed us to directly evaluate how well our approach reproduces the known physical parameters.

First, we conducted this test under two configurations: systems with total \NHI\ $<10^{17.2}$~\sqcm\ (optically thin) and those with total \NHI\ $>10^{17.2}$~\sqcm\ (optically thick). Table~\ref{tab:toy_model} presents the first case, considering two systems: one with fixed metallicity and the other with fixed density; while allowing the remaining parameters to vary. Note that, in principle, these two parameters can vary simultaneously. However, to verify whether our code correctly recovers the true value of a parameter, we kept one of the parameters fixed. Here, the \NHI\ weighted values are calculated by weighting the metallicity or density of each component by its respective \HI\ column density, i.e., $\langle X \rangle_{N(\mathrm{HI})} = \frac{\sum_i X_i \, N_i(\mathrm{HI})}{\sum_i N_i(\mathrm{HI})} \, $, where $X_i$ denotes the metallicity or density of the $i$-th component and $N_i(\mathrm{HI})$ is the corresponding \HI\ column density. The ``model-predicted'' value corresponds to the result from our Bayesian modeling applied to the summed column density of all components. In both configurations, our method reliably recovers values close to the true fixed inputs (also close to the \NHI-weighted values), demonstrating the internal robustness of the modeling framework.  

We further considered an additional example in which both the metallicity and gas density were varied such that the associated metal column densities (e.g., \CIV) will lie well above the detection sensitivity of the quasar spectra. In this case as well, the model-predicted values agree with the \NHI-weighted values, consistent with the previous cases.


\begin{table}
    \centering
    \caption{Results for the optically thin cases}
    \begin{tabular}{cccc}
    \hline
    Example type & (\NHI, \met, $\log($\nh$/\rm cm^{-3})$) &  $N$(\HI)-weighted & Model predicted\\
    \hline
    Constant metallicity & $(14.8, -1, -4.0)$ & \met$=-1$, \nh$=-3.44$ & \met$=-1$, \nh$=-3.49$ \\
    & $(15.0, -1, -3.0)$    &  &  \\
    & $(16.0, -1, -3.5)$    &  &  \\
    \hline

    Constant density & $(14.0, -0.5, -3.5)$     & \met=-1.98, \nh=-3.5  & \met=-1.98, \nh=-3.45 \\
    & $(14.5, -1.0, -3.5)$     &  &    \\
    & $(14.8, -1.5, -3.5)$     &  &    \\
    & $(15.0, -2.0, -3.5)$     &  &    \\
    & $(16.0, -3.0, -3.5)$     &  &    \\
    \hline

    Varying both metallicity and density & $(14.0, -0.5, -2.5)$ & \met=-1.42, \nh=-3.26  & \met=-1.61, \nh=-3.20 \\
    & $(13.8, -1.5, -4.0)$     &  &    \\
    & $(15.0, -2.0, -3.5)$     &  &    \\
    & $(14.5, -2.0, -3.0)$     &  &    \\
    \hline
    
    \end{tabular}
    \label{tab:toy_model}
\end{table}

\begin{table}
\centering
\caption{Results for the optically thick cases}
\begin{tabular}{lccc}
\toprule 
    (\NHI, \met, $\log($\nh$/\rm cm^{-3})$) & $N$(\HI) weighted & Constrained from & Model predicted\\
    \hline
     $(14.8, -1, -4.0)$ & \met$= -1$, \nh$=-3.5$  & high-ions & \met$=-1$, \nh$=-3.51$ \\
    $(15.0, -1, -3.0)$   &  &  &  \\
    \cmidrule(lr){3-4}
    $(16.0, -1, -3.5)$   &  & low-ions & \met$=-1.09$, \nh$=-3.35$\\
    $(17.5, -1, -3.5)$  &  &  &  \\
    
    \hline
    
    $(14.8, -1, -4.0)$  & \met$= -1$, \nh$=-1.5$  & high-ions & \met$=-3.3$, \nh$=-4.03$ \\
    $(15.0, -1, -3.0)$  &  &  &  \\
    \cmidrule(lr){3-4}
    $(16.0, -1, -3.5)$   &   & low-ions & \met$=-1.02$, \nh$=-1.58$ \\
    $(17.5, -1, -1.5)$  &  &  &  \\
    \hline
    \end{tabular}
    \\ Note: `Constrained from' indicates the set of ionic column densities used in the modeling, i.e., high-ions (e.g., \CIV, \SiIV) or low-ions (e.g., \SiII, \AlII).
    \label{tab:toy_model_gtr17}
\end{table}

The next case concerns systems with total \NHI\ exceeding $10^{17}$~\sqcm, corresponding to the Lyman limit systems. This scenario is particularly noteworthy because, as mentioned earlier, \CIV\ is no longer the dominant tracer. As shown in Table~\ref{tab:toy_model_gtr17}, the gas density and metallicity can be reliably constrained using low-ion transitions (e.g., \SiII, \AlII). Whereas, high-ion transitions (e.g., \CIV, \SiIV) only give reliable results in the low-density regime. Hence, if low-ions remain detectable in such high-\NHI\ systems, they should be used in the modeling. 

As a natural next step, we plan to perform detailed cloud-by-cloud modeling of the absorbers, following the approach of \citet{sameer2021}. This will allow us to derive the metallicity, density, and temperature of individual components, rather than the component-averaged quantities used in this work. With these component-level estimates in hand, we will be able to compute additional physical parameters, such as cloud sizes that cannot be reliably determined at present due to the use of averaged \nh\ and \met\ values for multi-component systems.

\section{The only system with a non-detection of \HI\ In our sample}
\label{sec:Q1317_lowlim} 
\begin{figure}
    \centering
    \hbox{
    \includegraphics[width=0.50\linewidth]{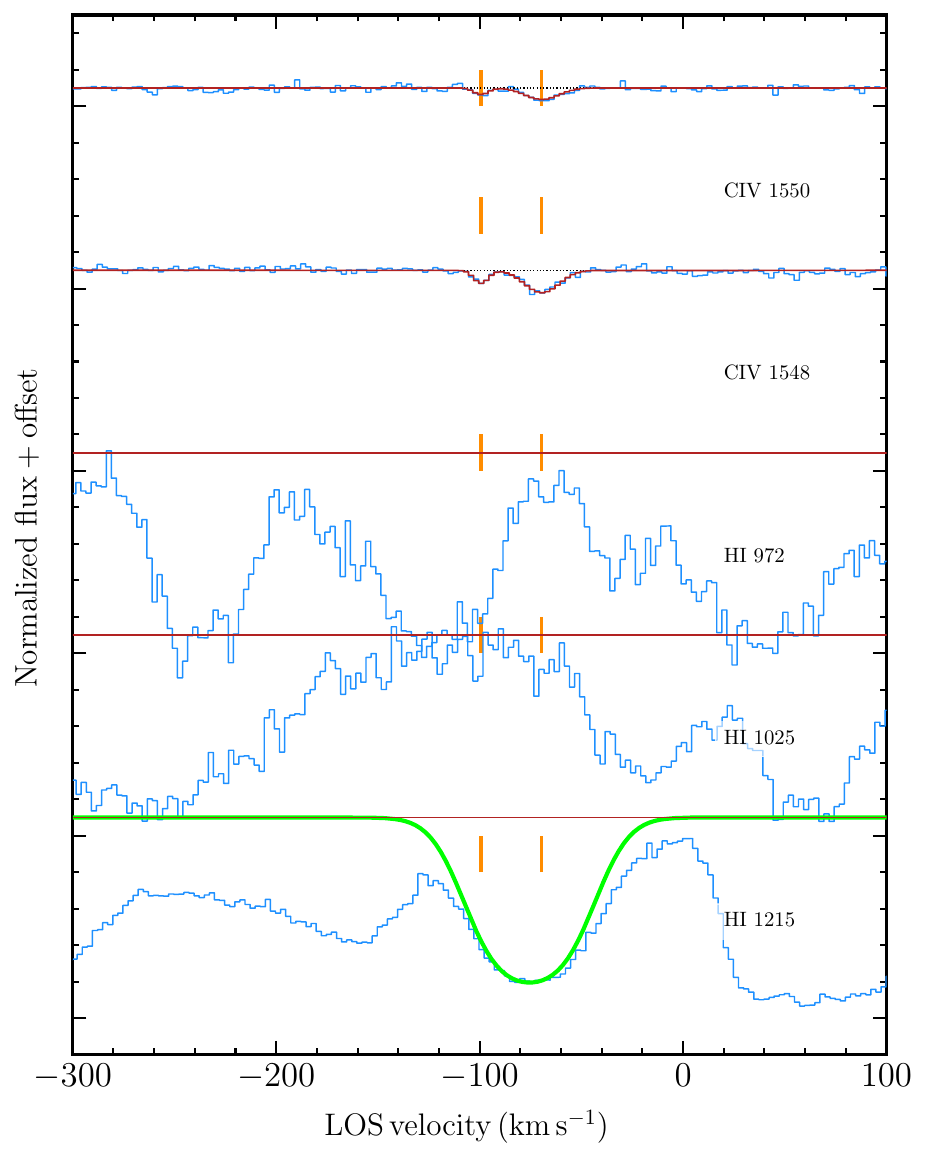}
    \includegraphics[width=0.49\linewidth]{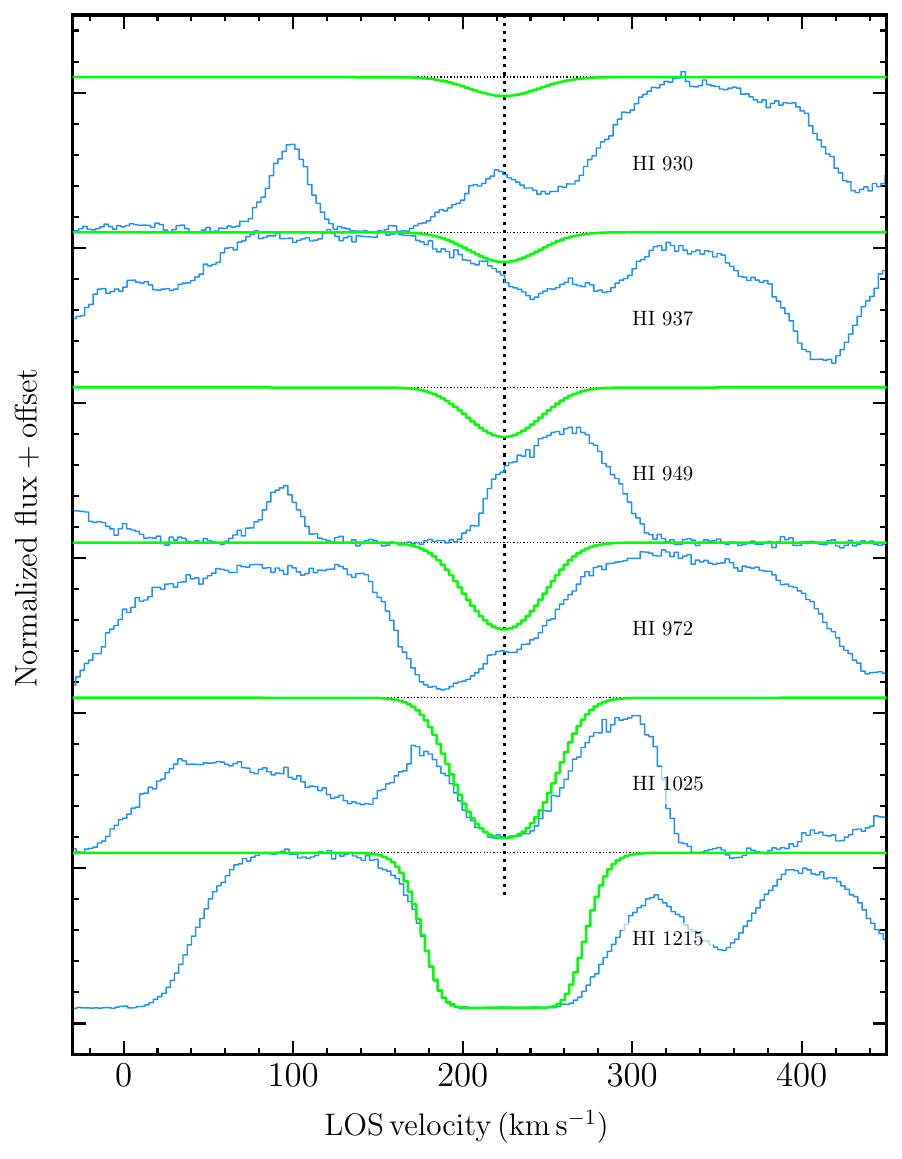}
    }
    \caption{{\tt Left:} Velocity plot of the absorber at $z_{\rm sys} \approx 2.9238$ (Q1317$-$0507 sightline), the only system in our sample where \HI\ is a non-detection giving rise to a lower-limit on the metallicity for this system. The quasar spectrum is in blue, and the best-fit Voigt profiles are shown in maroon. The $0$~\kms\ indicates the galaxy redshift. Vertical orange ticks mark the positions of \CIV\ components (in all the panels). No \HI\ is detected for this system, once the contribution of \lyb\ absorption from the $z=3.6458$ is factored in. The green absorption feature at the similar location as that of \CIV\ absorption is actually \HI$-$1025 from a higher-redshift system at $z_{\rm sys} \approx 3.6458$.
    {\tt Right:} Velocity profiles of this higher-$z$ \HI\ component (marked by the vertical black dotted line) with higher order Lyman series lines that shows \NHI\ for this component is well constrained. }
    \label{fig:Q1317_lowlim}
\end{figure}

{In this section, we discuss the only absorption system in our sample detected along the Q1317$-$0507 sightline, for which we have estimated an upper limit of \NHI, resulting in a lower limit on the metallicity.} The left panel of Fig.~\ref{fig:Q1317_lowlim} displays the absorption line profiles for this system at $z_{\rm sys} \approx 2.9238$. \CIV\ is the only metal transition detected. The maroon curves indicate the best-fitting Voigt profiles for the detected transitions.

Interestingly, this system lacks a detectable \HI\ counterpart. The green absorption feature in the left panel, although located near the expected position of \HI$\lambda$1215, actually corresponds to a \HI$\lambda$1025 line from a higher-redshift system at $z_{\rm sys} \approx 3.6458$. The velocity plot of the Lyman series lines for this higher-$z$ absorber is shown in the right panel of Fig.~\ref{fig:Q1317_lowlim}. Note that, the \NHI\ for the higher-$z$ absorber is nicely constrained using the higher order Lyman series lines.

To model the $z \approx 2.9238$ system, we first normalized the quasar spectrum at this location using the best-fitting Voigt profile of the higher-redshift absorber. We then derived an upper limit of $\log\, N({\rm HI})/{\rm cm^{-2}} \lesssim 12.8$. When we performed photoionization modeling using this upper limit, the posterior distribution of the metallicity pushed against the maximum value allowed by our Cloudy grid, i.e., \met~$=-0.09^{+0.47}_{-0.16}$, implying that this is a strict lower limit. The hydrogen number density is $\log\,(n_{\rm H}/\rm cm^{-3})= -3.55^{+0.33}_{-0.56}$, determined using a Gaussian prior (see Section~\ref{sec:Photoionization_model} for details of prior selection).
%


\bsp	
\label{lastpage}
\end{document}